\documentclass[12pt,preprint]{elsarticle}
\usepackage[latin9]{inputenc}
\usepackage{amsmath}
\usepackage{amssymb}

\makeatletter
\usepackage{tikz}

\usepackage{color}
\usepackage[mathlines]{lineno}

\numberwithin{equation}{section}

\newcommand{\nodeiiarb}[2]{$\begin{array}{c} {#1} \\ \small {#2} \end{array}$}
\newcommand{\nodeii}[2]{\nodeiiarb{#1}{({#2})}}

\journal{}

\makeatother

\begin{document}
\begin{frontmatter}

\title{Group classification of the two-dimensional magnetogasdynamics equations
in Lagrangian coordinates}

\author[SUTaddress]{S.V.~Meleshko \corref{cor1}}

\cortext[cor1]{Corresponding author}


\ead{sergey@math.sut.ac.th}

\author[SUTaddress]{E.I.~Kaptsov}

\ead{evgkaptsov@gmail.com}

\author[SUaddress]{S.~Moyo}

\ead{smoyo@sun.ac.za}

\author[UAHaddress]{G.M. ~Webb}

\ead{gmw0002@uah.edu}

\address[SUTaddress]{School of Mathematics, Institute of Science, Suranaree University
of Technology, 30000, Thailand}

\address[SUaddress]{Research, Innovation and Postgraduate Studies, Stellenbosch University,
South Africa}

\address[UAHaddress]{Center for Space Plasma and Aeronomic Research, The University of
Alabama in Huntsville, Huntsville, \\
 AL 35805, USA}
\begin{abstract}
The present paper is devoted to the group classification of magnetogasdynamics
equations in which dependent variables in Euler coordinates depend
on time and two spatial coordinates. It is assumed that the continuum
is inviscid and nonthermal polytropic gas with infinite electrical
conductivity. The equations are considered in mass Lagrangian coordinates.
Use of Lagrangian coordinates allows reducing number of dependent
variables. The analysis presented in this article gives complete group
classification of the studied equations. This analysis is necessary
for constructing invariant solutions and conservation laws on the
base of Noether's theorem.
\end{abstract}
\end{frontmatter}

\section{Introduction}

The equations of magnetogasdynamics (MGD) describe motion of a gas
under the action of the internal forces, which consist of the pressure
and magnetic forces. These equations describe phenomena related to
plasma flows, for example, in plasma confinement, as well as physical
problems in astrophysics and fluid metals flows.

The present article considers MGD flows in which dependent variables
in Euler coordinates depend on time and two spatial coordinates. It
is assumed that the continuum is inviscid and non-thermal polytropic
gas with infinite electrical conductivity. For the analysis of equations
describing the behavior of such a continuum, the Lie group analysis
method is applied.

Lie point symmetries are an effective tool for analyzing nonlinear
differential equations \cite{bk:Ovsiannikov1962,bk:Ovsiannikov1978,bk:Ibragimov[1983],bk:Olver[1986],bk:BlumanKumei1989}.
They are related with the fundamental physical principles of the model
under consideration and correspond to the important properties of
the differential equations. Finding an admitted Lie group is one of
the first and necessary steps in application of the group analysis
method to partial differential equations. Using found symmetries one
can construct a representation of invariant or partially invariant
solution. The representation of a solution reduces the number of the
independent variables. The group analysis method guarantees that the
reduced system of equations for an invariant solution has fewer independent
variables and is involutive. Admitted symmetry of variational partial
differential equations is a necessary condition for application of
Noether's theorem, which is used for deriving conservation laws.

Applications of the group analysis method for different versions of
MGD equations have been considered in many publications. For example,
the case of the finite conductivity was investigated in \cite{art:Gridnev1968,art:DorMHDpreprint1976}.
The case of the infinite conductivity was examined in \cite{bk:HandbookLie,art:PaliathanasisMHD2021}.
Invariant solutions were considered in \cite{art:OliveriSpeciale2005,art:Picard2008,art:Golovin_2009,bk:Golovin2011,%
art:GolovinSesma_2019,bk:Webb2018,art:WebbAnco2019}.
Comprehensive analysis of MGD equations in Eulerian and Lagrangian
coordinates with plain and cylindrical symmetries were given in \cite{art:DorKapKozMelMuk,art:DorKapKozMel2023}.

The present paper is devoted to the group classification of the MGD
equations, where all dependent variables in Eulerian coordinates depend
on time and two spatial coordinates\footnote{Such solutions can also be three-dimensional.}.
The study is performed in mass Lagrangian coordinates. The transition
to mass Lagrangian coordinates makes it possible to solve four MGD
equations. As a result of this solving, four arbitrary functions of
the mass Lagrangian coordinates are obtained. In group analysis, these
functions are called arbitrary elements. The presence of arbitrary
elements requires a group classification, which consists of finding
all Lie groups admitted by a system of partial differential equations
\cite{bk:Ovsiannikov1978,bk:Ibragimov[1983],bk:Olver[1986]}.
In practice, groups are represented by their generators. The generators
admitted for all arbitrary elements form the kernel of the admitted
Lie algebras. The group classification represents all non-equivalent
extensions of the kernel and the corresponding concrete forms of arbitrary
elements, where the equivalence is considered with respect to equivalence
transformations that preserve the structure of the equations, but
can change arbitrary elements.

The paper is organized as follows. The next section provides MGD equations
in mass Lagrangian coordinates. Derivation of the equations in Lagrangian
coordinates, when the dependent variables in Eulerian coordinates
depend on time and two independent space variables. Section \ref{equiv}
provides equivalence transformations, which are used for simplification
arbitrary elements. Sections \ref{nonisentr_b01_neq0} and \ref{isentr_b01_neq0}
give the group classifications of nonisentropic and isentropic solutions
when $b_{01}^{2}+b_{02}^{2}\neq0$. Sections \ref{nonisentr_b01_eq0}
and \ref{isentr_b01_eq0} are devoted to the group classifications
of nonisentropic and isentropic solutions when $b_{01}^{2}+b_{02}^{2}=0$.
Conclusions are stated in the last Section.

\section{Magnetogasdynamics equations in mass Lagrangian coordinates}

The magnetogasdynamics equations of an ideal perfect polytropic gas
can be written in the following form \cite{bk:KulikovskiiLubimov1965,bk:Webb2018}
\begin{equation}
\begin{array}{c}
D\rho+\rho\,div\,{\bf u}=0,\\
D{\bf u}+\rho^{-1}\nabla(p+\frac{1}{2}{\bf H}^{2})-\rho^{-1}({\bf H}\cdot\nabla){\bf H}=0,\\
D{\bf H}+{\bf H}\,div\,{\bf u}-({\bf H}\cdot\nabla){\bf u}=0,\,\,\,div\,{\bf H}=0,\\
DS=0,
\end{array}\label{eq:three_D_magnet}
\end{equation}
where $\rho$, ${\bf u}$, $p$, $S$, and ${\bf H}$ correspond to
the gas density, fluid velocity, pressure, entropy and magnetic induction,
respectively, and $\gamma$ is the polytropic exponent,
\[
D=\partial_{t}+{\bf u}\cdot\nabla,\,\,{\bf H}=(H_{1},H_{2},H_{3}),\,\,{\bf u}=(u_{1},u_{2},u_{3}),\,\,\,\boldsymbol{x}=(x_{1},x_{2},x_{3}).
\]
The magnetic field strength $\boldsymbol{{\bf H}}$ and magnetic field
induction ${\bf B}$ are related by the equation ${\bf B}=\sqrt{\mu_{0}}{\bf H}$,
where $\mu_{0}$ is the magnetic permeability. The pressure $p$,
the density $\rho$ and the entropy $S$ are related by the state
equation $p=A(S)\rho^{\gamma}$, where $A(S)=Re^{(S-S_{0})/c_{v}}$,
$R$ is the gas constant, $c_{v}$ is the dimensionless specific heat
capacity at constant volume, and $S_{0}$ is constant.

In coordinate form equations (\ref{eq:three_D_magnet}) become
\begin{subequations}
\label{eq:main}
\begin{gather}
\rho_{t}+u_{i}\rho_{x_{i}}+\rho u_{ix_{i}}=0,\label{eq:cons_mass}\\
\rho(u_{jt}+u_{i}u_{jx_{i}})+H_{i}H_{ix_{j}}-H_{i}H_{jx_{i}}+p_{x_{j}}=0,\,\,\,(j=1,2,3),\label{eq:momentum}\\
H_{jt}+u_{i}H_{jx_{i}}+H_{j}u_{ix_{i}}-H_{i}u_{jx_{i}}=0,\,\,\,(j=1,2,3),\label{Faraday's_law}\\
H_{ix_{i}}=0,\label{Gauss'_law}\\
S_{t}+u_{i}S_{x_{i}}=0,\label{eq:entropy}
\end{gather}
\end{subequations}
 where the energy equation is rewritten. Here summation with respect
to a repeated index is assumed.

The mass Lagrangian coordinates are introduced by the relations
\begin{equation}
\rho=J^{-1},\,\,\,\varphi_{it}(t,\xi)=u_{i}(t,\varphi(t,\xi)),\label{eq:nov07.1}
\end{equation}
where
\[
\xi=(\xi_{1},\xi_{2},\xi_{3}),\,\,\,\varphi=(\varphi_{1},\varphi_{2},\varphi_{3}),\,\,\,J=\det\left(\frac{\partial\varphi}{\partial\xi}\right),\,\,\,T=\frac{\partial\varphi}{\partial\xi}=\left(\begin{array}{ccc}
\varphi_{1,1} & \varphi_{2,1} & \varphi_{3,1}\\
\varphi_{1,2} & \varphi_{2,2} & \varphi_{3,2}\\
\varphi_{1,3} & \varphi_{2,3} & \varphi_{3,3}
\end{array}\right),
\]
and ${\displaystyle \varphi_{i,j}=\frac{\partial\varphi_{i}}{\partial x_{j}}}$.

In mass Lagrangian coordinates the conservation law of mass (\ref{eq:cons_mass})
becomes identical and equation (\ref{eq:entropy}) gives that $S=S_{0}(\xi)$,
where $S_{0}(\xi)$ is an arbitrary function.

For the sake of completeness we provide here the transition of equations
(\ref{eq:main}) to mass Lagrangian coordinates \citep{bk:Webb2018}.
Let $A=JT^{-1}$, then
\[
A_{ik}T_{kl}=J\delta_{il}.
\]
Noting that ${\displaystyle \frac{\partial}{\partial\xi_{j}}=\varphi_{i,j}\frac{\partial}{\partial x_{i}}}$,
the operators
\[
\frac{\partial}{\partial x}=\left(\begin{array}{c}
\frac{\partial}{\partial x_{1}}\\
\frac{\partial}{\partial x_{2}}\\
\frac{\partial}{\partial x_{3}}
\end{array}\right),\,\,\,\frac{\partial}{\partial\xi}=\left(\begin{array}{c}
\frac{\partial}{\partial\xi_{1}}\\
\frac{\partial}{\partial\xi_{2}}\\
\frac{\partial}{\partial\xi_{3}}
\end{array}\right),
\]
can be represented as follows
\[
\frac{\partial}{\partial\xi}=T\frac{\partial}{\partial x},\,\,\,\frac{\partial}{\partial x}=J^{-1}A\frac{\partial}{\partial\xi}.
\]

Gauss' law (\ref{Gauss'_law}) gives
\begin{equation}
JH_{ix_{i}}=A_{ik}H_{i\xi_{k}}=0.\label{eq:Gauss' law_L}
\end{equation}

Direct calculations show that
\begin{equation}
\frac{\partial}{\partial\xi_{k}}(A_{ik})=0,\,\,\,\forall i.\label{eq:prop1}
\end{equation}
The latter leads to the relations
\[
A_{ik}H_{j\xi_{k}}=(A_{ik}H_{j})_{\xi_{k}},\,\,\,\,A_{ik}p_{\xi_{k}}=(A_{ik}p)_{\xi_{k}},\,\,\,\forall i,\,j.
\]
Using these relations, the part of momentum equations (\ref{eq:momentum})
in Lagrangian coordinates become
\[
H_{i}H_{ix_{j}}-H_{i}H_{jx_{i}}+p_{x_{j}}=J^{-1}(H_{i}A_{jk}H_{i\xi{}_{k}}-H_{i}A_{ik}H_{j\xi_{k}}+A_{jk}p_{\xi_{k}})=
\]
\[
=J^{-1}(\frac{1}{2}A_{jk}\frac{\partial H^{2}}{\partial\xi{}_{k}}-H_{i}(A_{ik}H_{j})_{\xi_{k}}+A_{jk}p_{\xi_{k}})=
\]
\[
=J^{-1}(\frac{\partial}{\partial\xi{}_{k}}(\frac{1}{2}A_{jk}H^{2}+A_{jk}p)-H_{i}(A_{ik}H_{j})_{\xi_{k}}).
\]
By virtue of Gauss' law (\ref{eq:Gauss' law_L}), one derives that
\[
H_{i}(A_{ik}H_{j})_{\xi_{k}}=(H_{i}A_{ik}H_{j})_{\xi_{k}}-A_{ik}H_{i\xi_{k}}H_{j}=(H_{i}A_{ik}H_{j})_{\xi_{k}}.
\]
Hence,
\[
H_{i}H_{ix_{j}}-H_{i}H_{jx_{i}}+p_{x_{j}}=J^{-1}\frac{\partial}{\partial\xi{}_{k}}\left(A_{jk}(\frac{1}{2}H^{2}+p)-H_{i}A_{ik}H_{j}\right)=
\]
\[
=J^{-1}\frac{\partial}{\partial\xi{}_{k}}\left(\delta_{ij}A_{ik}(\frac{1}{2}H^{2}+p)-H_{i}A_{ik}H_{j}\right)=
\]
\[
=J^{-1}\frac{\partial}{\partial\xi{}_{k}}\left(A_{ik}\left(\delta_{ij}(\frac{1}{2}H^{2}+p)-H_{i}H_{j}\right)\right)
\]
Then the momentum equations in Lagrangian coordinates have the form
\[
\frac{\partial^{2}\varphi_{j}}{\partial t^{2}}+\frac{\partial}{\partial\xi{}_{k}}\left(A_{ik}\left(\delta_{ij}(\frac{1}{2}H^{2}+p)-H_{i}H_{j}\right)\right)=0.
\]

Faraday's equations (\ref{Faraday's_law}) in Lagrangian coordinates
reduces as follows. Let $\boldsymbol{\boldsymbol{b}}=\rho^{-1}\boldsymbol{{\bf H}}$,
then using the conservation law of mass and Faraday's equations, one
obtains
\[
\frac{db_{j}}{dt}=-\rho^{-2}\frac{d\rho}{dt}H_{j}+\rho^{-1}\frac{dH_{j}}{dt}=\rho^{-1}H_{i}u_{jx_{i}}=b_{i}u_{jx_{i}}.
\]
Introducing the vector $\boldsymbol{\boldsymbol{b}}_{0}$ such that
$\boldsymbol{\boldsymbol{b}}=T\boldsymbol{\boldsymbol{b}}_{0}$, one
derives
\[
\frac{\partial b_{j}}{\partial t}-J^{-1}b_{i}A_{ik}u_{j\xi_{k}}=\frac{\partial b_{0\alpha}}{\partial t}\varphi_{j,\alpha}+b_{0\alpha}u_{j\xi_{\alpha}}-b_{0\alpha}(J^{-1}T_{\alpha i}A_{ik})u_{j\xi_{k}}=\frac{\partial b_{0\alpha}}{\partial t}\varphi_{j,\alpha}=0.
\]
The latter gives that
\[
\frac{\partial b_{0\alpha}}{\partial t}=0,\,\,\,\forall\alpha.
\]
Hence, similar to the entropy, one integrates the Faraday's equation
$\boldsymbol{\boldsymbol{b}}_{0}=\boldsymbol{\boldsymbol{b}}_{0}(\xi)$,
where $\boldsymbol{\boldsymbol{b}}_{0}(\xi)=(b_{01}(\xi),b_{02}(\xi),b_{03}(\xi))$
is an arbitrary vector function of $\xi$. Gauss's equation (\ref{eq:Gauss' law_L})
reduces as follows
\[
H_{ix_{i}}=(\rho b_{i})_{x_{i}}=(\rho b_{0\alpha}\varphi_{i,\alpha})_{x_{i}}=J^{-1}A_{ik}(\rho b_{0\alpha}\varphi_{i,\alpha})_{\xi_{k}}=J^{-1}(J^{-1}A_{ik}b_{0\alpha}\varphi_{i,\alpha})_{\xi_{k}}
\]
\[
=J^{-1}(J^{-1}T_{\alpha i}A_{ik}b_{0\alpha})_{\xi_{k}}=J^{-1}(b_{0k})_{\xi_{k}}=0.
\]
Therefore, in mass Lagrangian coordinates equations (\ref{eq:main})
reduce to the equations
\begin{equation}
\frac{\partial^{2}\varphi_{j}}{\partial t^{2}}+\frac{\partial}{\partial\xi{}_{k}}\left(A_{ik}\left(\delta_{ij}(\frac{1}{2}H^{2}+p)-H_{i}H_{j}\right)\right)=0,\,\,\,(j=1,2,3),\,\,\,\frac{\partial}{\partial\xi{}_{k}}b_{0k}=0,\label{eq:3Dequations_inL}
\end{equation}
\[
S=S(\xi),\,\,\,\boldsymbol{b}_{0}=\boldsymbol{b}_{0}(\xi),
\]
where
\[
H_{i}=J^{-1}b_{0\alpha}\varphi_{i,\alpha},\,\,\,H^{2}=J^{-2}b_{0\alpha}b_{0\beta}\varphi_{i,\alpha}\varphi_{i,\beta}.
\]

\section{Equations (\ref{eq:three_D_magnet}) with two independent space variables
in Lagrangian coordinates}

We study the case, where all dependent functions in Eulerian coordinates
only depend on two space variables $x_{1}$ and $x_{2}$. From equations
(\ref{eq:nov07.1}) one obtains the Cauchy problem\footnote{Here the Lagrangian space variables $\xi_{i}$ are considered before
the transition to the mass Lagrangian coordinates.}

\[
(\varphi_{1,3})_{t}=u_{1x_{1}}\varphi_{1,3}+u_{1x_{2}}\varphi_{2,3},\,\,\,\varphi_{1,3}(0,\xi_{1},\xi_{2},\xi_{3})=0,
\]
\[
(\varphi_{2,3})_{t}=u_{2x_{1}}\varphi_{1,3}+u_{2x_{2}}\varphi_{2,3},\,\,\,\varphi_{2,3}(0,\xi_{1},\xi_{2},\xi_{3})=0.
\]
For sufficiently smooth functions $\boldsymbol{u}(t,\boldsymbol{x})$
the latter Cauchy problem has unique solution $\varphi_{i,3}=0,\,\,\,(i=1,2)$
that means
\[
\varphi_{1}=\varphi(t,\xi_{1},\xi_{2}),\,\,\,\varphi_{2}=\zeta(t,\xi_{1},\xi_{2}).
\]
In this case the transition from Lagrangian coordinates to the mass
Lagrangian coordinates can be done such that $\varphi_{3}(0,\xi_{1},\xi_{2},\xi_{3})=\xi_{3}$.
Hence, because of the uniqueness of a solution of the Cauchy problem
\[
(\varphi_{3,3})_{t}=u_{3x_{1}}\varphi_{\xi_{3}}+u_{3x_{2}}\zeta_{\xi_{3}}=0,\,\,\,\varphi_{3,3}(0,\xi_{1},\xi_{2},\xi_{3})=1,
\]
on gets $\varphi_{3}=\xi_{3}+\chi(t,\xi_{1},\xi_{2})$. Further we
use the notations $\,\xi_{1}=\xi,\,\,\,\xi_{2}=\eta$. Thus, one has
\[
T=\frac{\partial\varphi}{\partial\xi}=\left(\begin{array}{ccc}
\varphi_{\xi} & \zeta_{\xi} & \chi_{\xi}\\
\varphi_{\eta} & \zeta_{\eta} & \chi_{\eta}\\
0 & 0 & 1
\end{array}\right),\,\,\,A=\left(-\begin{array}{ccc}
\zeta_{\eta} & -\zeta_{\xi} & \chi_{\eta}\zeta_{\xi}-\chi_{\xi}\zeta_{\eta}\\
\varphi_{\eta} & \varphi_{\xi} & \chi_{\xi}\varphi_{\eta}-\chi_{\eta}\varphi_{\xi}\\
0 & 0 & \varphi_{\xi}\zeta_{\eta}-\varphi_{\eta}\zeta_{\xi}
\end{array}\right),\,\,\,J=\varphi_{\xi}\zeta_{\eta}-\varphi_{\eta}\zeta_{\xi},
\]
\[
b_{1}=b_{01}\varphi_{\xi}+b_{02}\varphi_{\eta},\,\,\,b_{2}=b_{01}\zeta_{\xi}+b_{02}\zeta_{\eta},
\]

\[
b_{3}=b_{01}\chi_{\xi}+b_{02}\chi_{\eta}+b_{03}.
\]
The latter relations provide that $b_{0i}=b_{0i}(\xi,\eta),\,\,\,(i=1,2,3)$.
As all functions only depend on $\xi$ and $\eta$, and the coefficients
$A_{31}=0$ and $A_{32}=0$, then equations (\ref{eq:3Dequations_inL})
become
\begin{subequations}
\label{eq:2Dequations}
\begin{gather}
{\displaystyle \frac{\partial^{2}\varphi_{j}}{\partial t^{2}}+\sum_{k=1}^{2}\sum_{i=1}^{2}\frac{\partial}{\partial\xi{}_{k}}\left(A_{ik}\left(\delta_{ij}(\frac{1}{2}H^{2}+p)-H_{i}H_{j}\right)\right)=0,\,\,\,(j=1,2),}\\
{\displaystyle \frac{\partial^{2}\chi}{\partial t^{2}}-\sum_{k=1}^{2}\sum_{i=1}^{2}\frac{\partial}{\partial\xi_{k}}\left(A_{ik}H_{i}H_{3}\right)=0,}\label{eq:3_direction}
\end{gather}
\end{subequations}
 where
\[
H_{1}=J^{-1}(b_{01}\varphi_{\xi}+b_{02}\varphi_{\eta}),\,\,\,H_{2}=J^{-1}(b_{01}\zeta_{\xi}+b_{02}\zeta_{\eta}),
\]
\[
H_{3}=J^{-1}(b_{01}\chi_{\xi}+b_{02}\chi_{\eta}+b_{03}),\,\,\,H^{2}=H_{1}^{2}+H_{2}^{2}+H_{3}^{2},
\]
and
\[
S=S(\xi,\eta),\,\,\,\boldsymbol{b}_{0}=(b_{01}(\xi,\eta),b_{02}(\xi,\eta),b_{03}(\xi,\eta))
\]
are arbitrary functions such that
\begin{equation}
\frac{\partial}{\partial\xi}b_{01}+\frac{\partial}{\partial\eta}b_{02}=0.\label{eq:nov08.1}
\end{equation}

\section{Equivalence transformations}

\label{equiv} The class of equations (\ref{eq:2Dequations}) is parameterized
by arbitrary elements $S(\xi,\eta)$, $b_{0i}(\xi,\eta)$, ($i=1,2,3$).
The first step of the group classification of the class of equations
of form (\ref{eq:2Dequations}) consists of describing the equivalence
among the equations of this class. The group classification is considered
with respect to these equivalence transformations.

Direct calculations show that the transformations corresponding to
the generators

\[
X_{1}^{e}=\frac{\partial}{\partial\xi},\,\,\,X_{2}^{e}=\frac{\partial}{\partial\eta},\,\,\,X_{3}^{e}=\frac{\partial}{\partial\varphi},\,\,\,X_{4}^{e}=\frac{\partial}{\partial\zeta},\,\,\,X_{5}^{e}=\frac{\partial}{\partial\chi},\,\,\,X_{6}^{e}=\frac{\partial}{\partial t},
\]
\[
X_{7}^{e}=t\frac{\partial}{\partial\varphi},\,\,\,X_{8}^{e}=t\frac{\partial}{\partial\zeta},\,\,\,X_{9}^{e}=t\frac{\partial}{\partial\chi},\,\,\,X_{10}^{e}=\zeta\frac{\partial}{\partial\varphi}-\varphi\frac{\partial}{\partial\zeta},
\]
\[
X_{11}^{e}=t\frac{\partial}{\partial t}+\xi\frac{\partial}{\partial\xi}+\eta\frac{\partial}{\partial\eta}+\varphi\frac{\partial}{\partial\varphi}+\zeta\frac{\partial}{\partial\zeta}+\chi\frac{\partial}{\partial\chi},
\]
\[
X_{12}^{e}=t\frac{\partial}{\partial t}+2\xi\frac{\partial}{\partial\xi}+2\eta\frac{\partial}{\partial\eta}+4(1-\gamma)S\frac{\partial}{\partial S}-2b_{03}\frac{\partial}{\partial b_{03}},
\]
\[
X_{13}^{e}=-t\frac{\partial}{\partial t}+2S\frac{\partial}{\partial S}+b_{01}\frac{\partial}{\partial b_{01}}+b_{02}\frac{\partial}{\partial b_{02}}+b_{03}\frac{\partial}{\partial b_{03}},
\]
\[
X_{\lambda_{1}}^{e}=\lambda_{1}(\xi,\eta)\frac{\partial}{\partial\chi},\,\,\,X_{\lambda_{2}}^{e}=t\lambda_{2}(\xi,\eta)\frac{\partial}{\partial\chi}.
\]
do not change the structure of equations (\ref{eq:2Dequations}) and
(\ref{eq:nov08.1}). Here the generators $X_{i}^{e},\,\,\,(i=3,4,...11)$
are inherited by equations in Eulerian coordinates (\ref{eq:main}),
where $X_{3}^{e}$, $X_{4}^{e}$, $X_{5}^{e}$ correspond to the shifts
with respect to $x_{i},\,\,\,(i=1,2,3)$; $X_{6}^{e}$, $X_{7}^{e}$,
$X_{8}^{e}$ correspond to the Galilean boosts; $X_{10}^{e}$ correspond
to the rotation. The generator $X_{\lambda_{1}}^{e}$ allows adding
a function $\lambda_{1}(\xi,\eta)$ to $\chi$. In particular, for
given $b_{0i}(\xi,\eta),\,\,\,(i=1,2,3)$ such that $b_{01}^{2}+b_{02}^{2}\neq0$,
choosing a function $\lambda_{1}(\xi,\eta)$ satisfying the condition
\[
b_{01}\lambda_{1\xi}+b_{02}\lambda_{1\eta}+b_{03}=0,
\]
one can assume that after the transformation $b_{03}=0$. Indeed,
for $\chi=\tilde{\chi}+\lambda_{1}$ one derives that
\[
b_{3}=b_{01}\tilde{\chi}_{\xi}+b_{02}\tilde{\chi}_{\eta}+b_{01}\lambda_{1\xi}+b_{02}\lambda_{1\eta}+b_{03}=b_{01}\tilde{\chi}_{\xi}+b_{02}\tilde{\chi}_{\eta}.
\]

There are also two involutions
\[
\begin{array}{c}
E_{1}:\,\,\,t\rightarrow-t,\\
E_{2}:\,\,\,(\xi,\eta,\varphi,\zeta,\chi)\rightarrow-(\xi,\eta,\varphi,\zeta,\chi),
\end{array}
\]
where only changeable variables are presented.

The admitted generator $X$ is sought in the form
\[
X=\xi^{\xi}\frac{\partial}{\partial\xi}+\xi^{\eta}\frac{\partial}{\partial\eta}+\xi^{t}\frac{\partial}{\partial t}+\zeta^{\varphi}\frac{\partial}{\partial\varphi}+\zeta^{\zeta}\frac{\partial}{\partial\zeta}+\zeta^{\chi}\frac{\partial}{\partial\chi},
\]
where all coefficients of the generator $X$ depend on $(t,\xi,\eta,\varphi,\zeta,\chi)$.
The determining equations \citep{bk:Ovsiannikov1978} are obtained
by applying the prolongation of the generator $X$ to the left-hand
side of equations (\ref{eq:2Dequations}):
\[
XF_{|(\ref{eq:2Dequations})}=0,
\]
where $F$ is the left-hand side of equations (\ref{eq:2Dequations}),
and $|(\ref{eq:2Dequations})$ means to consider $XF$ on the manifold
defined by equations (\ref{eq:2Dequations}).

The analysis of the determining equations depend on the relations
between the entropy $S(\xi,\eta)$ and the vector $\boldsymbol{b}_{0}(\xi,\eta)$.
It breaks down into several cases. Globally, according to the equivalence
transformations corresponding to the generator $X_{f}^{e}$, it decomposes
into $b_{01}^{2}+b_{02}^{2}\neq0$ and $b_{01}^{2}+b_{02}^{2}=0$,
and each of these cases is divided into non-isentropic and isentropic
solutions.

\section{Nonisentropic case with $b_{01}^{2}+b_{02}^{2}\protect\neq0$}

\label{nonisentr_b01_neq0}

The general solution of Gauss' equation (\ref{eq:nov08.1}) can be
written as
\[
b_{01}=\psi_{\eta},\,\,\,b_{02}=-\psi_{\xi},
\]
where $\psi=\psi(\xi,\eta)$. One also can assume that $\psi_{\eta}\neq0$.
By virtue of the equivalence transformation corresponding to the generator
$X_{f}^{e}$ it can be considered that $b_{03}=0$.

Partially solving the determining equations one derives that $\xi^{\xi}=\xi^{\xi}(\xi,\eta)$,
$\xi^{\eta}=\xi^{\eta}(\xi,\eta)$, and

\[
\zeta^{\varphi}=(k_{2}+\frac{1}{2}k_{5})\varphi+k_{1}\zeta+k_{7}t+k_{8},\,\,\,\zeta^{\zeta}=(k_{2}+\frac{1}{2}k_{5})\zeta-k_{1}\varphi+k_{9}t+k_{10},
\]
\[
\zeta^{\chi}=(k_{2}+\frac{1}{2}k_{5})\chi+t\lambda_{1}+\lambda_{2},\,\,\,\xi^{t}=k_{5}t+k_{6},
\]
where $k_{i}$ are constant, and $\lambda_{i}=\lambda_{i}(\psi)$
are arbitrary functions. The remaining equations are

\begin{equation}
\begin{array}{c}
2(1-\gamma)\xi_{\eta}^{\xi}\psi_{\eta}Sj_{1}+\xi^{\xi}\left(2S(1-\gamma)(j_{1\eta}\psi_{\eta}+g(S_{\eta\eta}\psi_{\eta}-\psi_{\eta\eta}S_{\eta})-(2\gamma-1)\psi_{\eta}S_{\xi}S_{\eta}\right)\\
+\xi^{\eta}\left(2(1-\gamma)S(S_{\eta\eta}\psi_{\eta}-\psi_{\eta\eta}S_{\eta})+(2\gamma-1)\psi_{\eta}S_{\eta}{}^{2}\right)+(2k_{2}(1-2\gamma)+k_{5})S\psi_{\eta}S_{\eta}=0,
\end{array}\label{eq:oct15.1}
\end{equation}
\begin{equation}
\begin{array}{c}
\xi^{\xi}\left(2(1-\gamma)S(j_{1}g\psi_{\eta\eta}+\psi_{\eta}j_{1\xi})+j_{1}\psi_{\eta}((2\gamma-1)S_{\xi}-2(\gamma-1)Sg_{\eta})\right)\\
+\xi^{\eta}\left(2(1-\gamma)S(\psi_{\eta\eta}j_{1}+j_{1\eta}\psi_{\eta})+(2\gamma-1)S_{\eta}\psi_{\eta}j_{1}\right)-(2k_{2}(\gamma+2)+k_{5}(2\gamma+3)Sj_{1}\psi_{\eta}=0,
\end{array}\label{eq:oct15.2}
\end{equation}
\begin{equation}
\begin{array}{c}
\xi_{\xi}^{\eta}=-g^{2}\xi_{\eta}^{\xi}-\xi^{\xi}\left(2gg_{\eta}+2\frac{\psi_{\eta\eta}}{\psi_{\eta}}g^{2}+g_{\xi}\right)\\
-\xi^{\eta}\left(g_{\eta}+2\frac{\psi_{\eta\eta}}{\psi_{\eta}}g\right)+(2k_{2}-k_{5})g,
\end{array}\label{eq:oct15.3}
\end{equation}
\begin{equation}
\begin{array}{c}
\xi_{\eta}^{\eta}=-g\xi_{\eta}^{\xi}-\xi^{\xi}\left(g\frac{\psi_{\eta\eta}}{\psi_{\eta}}+g_{\eta}+\frac{S_{\xi}}{2(\gamma-1)S}\right)\\
-\xi^{\eta}\left(\frac{\psi_{\eta\eta}}{\psi_{\eta}}+\frac{S_{\eta}}{2(\gamma-1)S}\right)+\frac{2(2\gamma-1)k_{2}-k_{5}}{2(\gamma-1)},
\end{array}\label{eq:oct15.4}
\end{equation}

\begin{equation}
\begin{array}{c}
\xi_{\xi}^{\xi}=g\xi_{\eta}^{\xi}+\xi^{\xi}\left(g_{\eta}+\frac{\psi_{\eta\eta}}{\psi_{\eta}}g-\frac{S_{\xi}}{2(\gamma-1)S}\right)\\
+\xi^{\eta}\left(\frac{\psi_{\eta\eta}}{\psi_{\eta}}-\frac{S_{\eta}}{2(\gamma-1)S}\right)-\frac{(2\gamma-3)k_{5}+2k_{2}}{2(\gamma-1)}.
\end{array}\label{eq:oct15.5}
\end{equation}
where
\begin{equation}
j_{1}=S_{\xi}-gS_{\eta},\,\,\,g=\frac{\psi_{\xi}}{\psi_{\eta}}.\label{eq:nov17.1}
\end{equation}
As a solution of equations (\ref{eq:oct15.1})-(\ref{eq:oct15.5})
determines an admitted Lie group of equations (\ref{eq:2Dequations}),
they are called the defining equations.

The generators admitted for any functions $S$, $b_{01}$ and $b_{02}$,
composes a Lie algebra, called the kernel of admitted Lie algebras.
A basis of this Lie algebra consists of the generators
\begin{equation}
\begin{array}{c}
{\displaystyle X_{1}=\frac{\partial}{\partial\varphi},\,\,\,X_{2}=\frac{\partial}{\partial\zeta},\,\,\,X_{3}=\frac{\partial}{\partial\chi},\,\,\,X_{4}=\frac{\partial}{\partial t},}\\
{\displaystyle X_{5}=t\frac{\partial}{\partial\varphi},\,\,\,X_{6}=t\frac{\partial}{\partial\zeta},\,\,\,X_{7}=\zeta\frac{\partial}{\partial\varphi}-\varphi\frac{\partial}{\partial\zeta},\,\,\,X_{\lambda_{1}}=\lambda_{1}\frac{\partial}{\partial\chi},\,\,\,X_{\lambda_{2}}=t\lambda_{2}\frac{\partial}{\partial\chi}.}
\end{array}\label{eq:kernel}
\end{equation}
The kernel extensions are discussed next.


\subsection{Case $j_{1}\protect\neq0$ }

Introducing
\[
h_{1}=\xi^{\xi}\psi_{\xi}+\xi^{\eta}\psi_{\eta},\,\,\,h_{2}=\xi^{\xi}S_{\xi}+\xi^{\eta}S_{\eta},
\]
one finds
\[
\xi^{\xi}=(\psi_{\eta}j_{1})^{-1}(-S_{\eta}h_{1}+\psi_{\eta}h_{2}),\ \ \xi^{\eta}=(\psi_{\eta}j_{1})^{-1}(S_{\xi}h_{1}-\psi_{\eta}gh_{2}).
\]
From equation (\ref{eq:oct15.4}) one obtains
\begin{equation}
h_{2}=S\left(2\frac{h_{1\eta}}{\psi_{\eta}}(1-\gamma)+2k_{2}(2\gamma-1)-k_{5}\right).\label{eq:oct8.2}
\end{equation}
Finding $h_{1\xi\eta}$ from equation (\ref{eq:oct15.5}), equation
(\ref{eq:oct15.3}) becomes
\begin{equation}
h_{1\xi}-h_{1\eta}g=0.\label{eq:oct8.1}
\end{equation}
Hence, $h_{1}=h_{1}(\psi)$, and equation (\ref{eq:oct15.5}) reduces
to
\begin{equation}
\begin{array}{c}
{\displaystyle h_{1\eta}j_{2}+h_{1}\left(\frac{(2\gamma+j_{2}-1)}{2(\gamma-1)}\frac{S_{\eta}}{S}-\frac{j_{1\eta}}{j_{1}}-\frac{\psi_{\eta\eta}}{\psi_{\eta}}\right)}\\
{\displaystyle +\frac{\psi_{\eta}}{\gamma-1}\left(k_{2}j_{2}(1-2\gamma)+\frac{1}{2}k_{5}j_{2}-k_{2}+\frac{3-2\gamma}{2}k_{5}\right)=0,}
\end{array}\label{eq:oct8.3}
\end{equation}
where
\begin{equation}
j_{2}=j_{1}^{-2}\left(2(\gamma-1)S(j_{1\xi}-gj_{1\eta}+j_{1}g_{\eta})-(2\gamma-1)j_{1}^{2}\right).\label{eq:oct8.4}
\end{equation}
Notice that from the notation (\ref{eq:oct8.4}) one has
\begin{equation}
j_{1\xi}=gj_{1\eta}-j_{1}g_{\eta}+\frac{j_{1}^{2}}{2(\gamma-1)S}(j_{2}+(2\gamma-1)).\label{eq:oct14.1}
\end{equation}


\subsubsection{Case $j_{2}\protect\neq0$}

From equation (\ref{eq:oct8.3}) one finds $h_{1\eta}$. Introducing
the function

\begin{equation}
j_{3}=j_{2\xi}-gj_{2\eta},\label{eq:nov17.2}
\end{equation}
the compatibility condition $(h_{1\xi})_{\eta}=(h_{1\eta})_{\xi}$
becomes
\begin{equation}
h_{1}\mu+2\psi_{\eta}^{2}Sj_{1}j_{3}\tilde{k}_{2}=0,\label{eq:oct8.6}
\end{equation}
where ${\displaystyle \tilde{k}_{2}=k_{2}+\frac{(2\gamma-3)}{2}k_{5}}$,
and
\begin{equation}
\mu=j_{3}\left(2(\gamma-1)S(j_{1\eta}\psi_{\eta}+\psi_{\eta\eta}j_{1})-S_{\eta}\psi_{\eta}j_{1}(2\gamma-1)\right)-j_{2\eta}\psi_{\eta}j_{1}^{2}j_{2}.\label{eq:oct8.7}
\end{equation}


Let $j_{3}\mu\neq0$. Introducing the function
\begin{equation}
j_{4}=-2\mu^{-1}\psi_{\eta}^{2}Sj_{1}j_{3},\label{eq:oct8.8}
\end{equation}
equation (\ref{eq:oct8.6}) gives that $h_{1}=j_{4}\tilde{k}_{2}$.
As for $\tilde{k}_{2}=0$ there is no an extension of the kernel of
admitted Lie algebras, and because $h_{1}=h_{1}(\psi)$, then
\[
j_{4}=j_{4}(\psi).
\]

From definition of $j_{4}$ one finds
\[
\psi_{\eta\eta}=\frac{\psi_{\eta}}{(\gamma-1)}\left(\frac{(2\gamma-1)S_{\eta}}{2S}-\frac{(\gamma-1)j_{1\eta}}{j_{1}}+\frac{j_{1}j_{2}j_{2\eta}}{2Sj_{3}}+\frac{\psi_{\eta}}{j_{4}}\right).
\]

The compatibility condition $(\psi_{\eta\eta})_{\xi}=(\psi_{\xi})_{\eta\eta}$
gives
\begin{equation}
j_{1\eta}=\frac{j_{1}}{j_{3}^{2}}(j_{3\eta}j_{2\xi}-j_{3\xi}j_{2\eta})+\frac{j_{1}S_{\eta}}{S}+\frac{j_{1j_{2\eta}}}{j_{3}}\left(\frac{j_{1}(j_{2}+1)}{2(\gamma-1)S}-g_{\eta}\right).\label{eq:condition_1}
\end{equation}
The relation $(j_{1\xi})_{\eta}=(j_{1\eta})_{\xi}$ provides the condition
\begin{equation}
\begin{array}{c}
{\displaystyle j_{3}^{2}(j_{2\eta}g_{\xi\eta}-j_{2\xi}g_{\eta\eta})+j_{3}(j_{5\xi}j_{2\eta}-j_{5\eta}j_{2\xi})}\\[2ex]
{\displaystyle +(j_{3}g_{\eta}+2j_{5})(j_{3\eta}j_{3}-j_{2\eta}j_{5})=0,}
\end{array}\label{eq:condition_2}
\end{equation}
where $j_{5}=j_{3\xi}-gj_{3\eta}$.

Substituting $h_{1}$ into (\ref{eq:oct8.3}) one derives
\begin{equation}
k_{5}=\tilde{k}_{2}\frac{1}{4(\gamma-1)^{2}}\left(2(\gamma-1)\frac{j_{4\eta}}{\psi_{\eta}}-j_{4}M+2(2\gamma-1)\right),\label{eq:oct8.10}
\end{equation}
where
\begin{equation}
M=\frac{S_{\eta}j_{3}-j_{2\eta}j_{1}}{\psi_{\eta}Sj_{3}}.\label{eq:def_mm}
\end{equation}
Direct calculations show that $M$ satisfies the relation
\[
M_{\xi}-gM_{\eta}=0,
\]
which means that $M=M(\psi)$.

For the existence of an extension of the kernel of admitted Lie algebras
one needs to assume that $k_{8}/\tilde{k}_{2}$ is constant. Thus,
\begin{equation}
2(\gamma-1)\frac{j_{4\eta}}{\psi_{\eta}}-j_{4}M=k,\label{eq:condition_3}
\end{equation}
where $k$ is some constant.

Equation (\ref{eq:oct15.1}) becomes
\begin{equation}
M_{\eta}=-\frac{g\psi_{\eta}M}{2(\gamma-1)}\left(M+\frac{k}{j_{4}}\right).\label{eq:condition_4}
\end{equation}

The extension of the kernel of admitted Lie algebras is defined by
the generator

\begin{equation}
\begin{array}{c}
X_{9}^{(1)}=\frac{j_{4}}{\psi_{\eta}j_{3}}\left(-j_{2\eta}\frac{\partial}{\partial\xi}+j_{2\xi}\frac{\partial}{\partial\eta}\right)-\frac{k+2(2\gamma-1)}{4(\gamma-1)^{2}}t\frac{\partial}{\partial t}\\
+\frac{k(\gamma-2)-2\gamma}{4(\gamma-1)^{2}}\left(\varphi\frac{\partial}{\partial\varphi}+\zeta\frac{\partial}{\partial\zeta}+\chi\frac{\partial}{\partial\chi}\right).
\end{array}\label{eq:oct8.12}
\end{equation}

Summarizing, one can state that if the functions $\psi(\xi,\eta)$
and $S(\xi,\eta)$ satisfy the conditions (\ref{eq:condition_1}),
(\ref{eq:condition_3}) and (\ref{eq:condition_4}), where $j_{i},\,\,\,(i=1,2,3,4)$
are defined by the formulas (\ref{eq:nov17.1}), (\ref{eq:oct8.4}),
(\ref{eq:nov17.2}) and (\ref{eq:oct8.8}), then the extension of
the kernel of admitted Lie algebras is defined by the generator (\ref{eq:oct8.12}).
Here condition (\ref{eq:condition_2}) guarantees the existence of
the functions $\psi(\xi,\eta)$, $S(\xi,\eta)$ satisfying conditions
(\ref{eq:condition_1}) and (\ref{eq:condition_3}). 

Case $j_{i}\neq0,\,\,\,(i=1,2,3)$ and $\mu=0$. Equation (\ref{eq:oct8.6})
provides that $\tilde{k}_{2}=0$. From $\mu=0$ one finds that

\begin{equation}
\psi_{\eta\eta}=-\frac{j_{1\eta}}{j_{1}}\psi_{\eta}+\frac{1}{2(\gamma-1)S}\left((2\gamma-1)S_{\eta}\psi_{\eta}+j_{2\eta}\psi_{\eta}j_{1}j_{2}j_{3}^{-1}\right).\label{eq:oct10.1}
\end{equation}
The compatibility relation $(\psi_{\eta\eta})_{\xi}=(\psi_{\xi})_{\eta\eta}$
is

\begin{equation}
\begin{array}{c}
{\displaystyle j_{1\eta}=\frac{j_{1}}{j_{3}^{2}}\left(j_{2\eta}\left(-g_{\eta}j_{3}-j_{5}+\frac{j_{1}j_{3}(j_{2}-3)}{2(\gamma-1)S}\right)+j_{3}\left(\frac{\gamma+1}{\gamma-1}\frac{S_{\eta}}{S}j_{3}+j_{3\eta}\right)\right)}-2\frac{j_{1}}{\gamma-1}\psi_{\eta}M\end{array},\label{eq:oct10.5}
\end{equation}
where $j_{5}=j_{3\xi}-gj_{3\eta}$. The compatibility condition $(j_{1\eta})_{\xi}=(j_{1\xi})_{\eta}$
also coincides with (\ref{eq:condition_2}).

Equation (\ref{eq:oct8.3}) becomes
\begin{equation}
\begin{array}{c}
{\displaystyle h_{1\eta}+\frac{M\psi_{\eta}}{2(\gamma-1)}h_{1}+2k_{5}(\gamma-1)\psi_{\eta}=0}.\end{array}\label{eq:oct10.2}
\end{equation}
Equation (\ref{eq:oct10.5}) provides that
\begin{equation}
M_{\xi}-gM_{\eta}=0,\label{eq:oct10.3}
\end{equation}
which also means that $M=M(\psi)$. Equation (\ref{eq:oct15.1}) reduces
to
\[
h_{1}\nu-4k_{5}(\gamma-1)^{2}M\psi_{\eta}=0,
\]
where $\nu=2(\gamma-1)M_{\eta}-M^{2}\psi_{\eta}$.

Assuming that $\nu\neq0$, one obtains
\[
h_{1}=k_{5}\lambda,
\]
where
\[
\lambda=\frac{4(\gamma-1)^{2}M\psi_{\eta}}{\nu}.
\]
For the existence of the extension of the kernel of admitted Lie algebras
it is necessary that $\lambda$ is constant, say $\lambda=k$:
\[
h_{1}=kk_{5}.
\]
Substituting the latter into (\ref{eq:oct10.2}),
\begin{equation}
M=-\frac{4(\gamma-1)^{2}}{k}\label{eq:oct10.2-1}
\end{equation}
or
\begin{equation}
\frac{S_{\eta}j_{3}-j_{2\eta}j_{1}}{\psi_{\eta}Sj_{3}}=-\frac{4(\gamma-1)^{2}}{k},\label{eq:condition_4_1a}
\end{equation}
and the extension of the kernel of admitted Lie algebras is defined
by the generator
\begin{equation}
X_{9}^{(2)}=\frac{k}{(\gamma-2)\psi_{\eta}j_{3}}\left(j_{2\eta}\frac{\partial}{\partial\xi}-j_{2\xi}\frac{\partial}{\partial\eta}\right)-\frac{1}{(\gamma-2)}t\frac{\partial}{\partial t}+\varphi\frac{\partial}{\partial\varphi}+\zeta\frac{\partial}{\partial\zeta}+\chi\frac{\partial}{\partial\chi}.\label{eq:oct12.2-1}
\end{equation}

Let $\nu=0$, then $k_{5}M=0$.

Consider $M=0$ or
\[
S_{\eta}j_{2\xi}-j_{2\eta}S_{\xi}=0.
\]
The latter means that $j_{2}=j_{2}(S)$. Integrating (\ref{eq:oct10.2}),
one obtains
\begin{equation}
h_{1}=-2k_{5}(\gamma-1)\psi+k_{12}.\label{eq:oct10.2-2}
\end{equation}
The extension of the kernel of admitted Lie algebras is defined by
the generators

\begin{equation}
X_{9}^{(3)}=\frac{2(\gamma-1)\psi}{(\gamma-2)\psi_{\eta}j_{3}}\left(-j_{2\eta}\frac{\partial}{\partial\xi}+j_{2\xi}\frac{\partial}{\partial\eta}\right)-\frac{1}{\gamma-2}t\frac{\partial}{\partial t}+\varphi\frac{\partial}{\partial\varphi}+\zeta\frac{\partial}{\partial\zeta}+\chi\frac{\partial}{\partial\chi}.\label{eq:oct12.2-1-1}
\end{equation}
\begin{equation}
X_{10}^{(3)}=\frac{1}{\psi_{\eta}j_{3}}\left(-j_{2\eta}\frac{\partial}{\partial\xi}+j_{2\xi}\frac{\partial}{\partial\eta}\right).\label{eq:oct12.2-1-1-1}
\end{equation}


If $M\neq0$, then $k_{5}=0$,
\[
M=-\frac{2(\gamma-1)}{\psi+k},
\]
with some constant $k$, and equation (\ref{eq:oct10.2}) reduces
to

\begin{equation}
\begin{array}{c}
{\displaystyle h_{1\eta}+\frac{M\psi_{\eta}}{2(\gamma-1)}h_{1}=0}.\end{array}\label{eq:oct10.2-3}
\end{equation}
Hence, $h_{1}=k_{12}/M$ and the extension of the kernel of admitted
Lie algebras is defined by the generator
\begin{equation}
X_{9}^{(4)}=\frac{1}{M\psi_{\eta}j_{3}}\left(-j_{2\eta}\frac{\partial}{\partial\xi}+j_{2\xi}\frac{\partial}{\partial\eta}\right).\label{eq:mov22.1}
\end{equation}

Case $j_{1}j_{2}\neq0$ and $j_{3}=0$. The assumption $j_{3}=0$
gives that $j_{2}=j_{2}(\psi)$, and equation (\ref{eq:oct8.6}) becomes
$h_{1}j_{2\eta}=0$. If $j_{2\eta}\neq0$, then $h_{1}=0$ and equation
(\ref{eq:oct8.3}) leads to the condition
\[
j_{2}\left(\tilde{k}_{2}(1-2\gamma)+2(\gamma-1)^{2}k_{5}\right)-\tilde{k}_{2}=0.
\]
As $j_{2\eta}\neq0$, then the latter equation provides that $\tilde{k}_{2}=0$
and $k_{5}=0$. Hence, for $j_{2\eta}\neq0$ there is no an extension
of the kernel of admitted Lie algebras. Thus, one should assume that
$j_{2\eta}=0$, which gives that $j_{2}=k$, where $k\neq0$ is constant.
Equation (\ref{eq:oct8.3}) reduces to
\begin{equation}
\begin{array}{c}
{\displaystyle h_{1\eta}-h_{1}\lambda\psi_{\eta}+\beta\psi_{\eta}=0},\end{array}\label{eq:oct12.5}
\end{equation}
where
\[
\lambda=\frac{1}{k}\left(\frac{\psi_{\eta\eta}}{\psi_{\eta}}+\frac{j_{1\eta}}{j_{1}}-\frac{(2\gamma+k-1)S_{\eta}}{2(\gamma-1)S}\right),\,\,\,\beta=2k_{5}(\gamma-1)-\tilde{k}_{2}\frac{(2\gamma-1)k+1}{k(\gamma-1)}.
\]
Finding $\psi_{\eta\eta}$ from the latter notation of $\lambda$,
the condition $(\psi_{\eta\eta})_{\xi}=(\psi_{\xi})_{\eta\eta}$ provides
that $\lambda=\lambda(\psi)$.

Equation (\ref{eq:oct15.1}) becomes
\begin{equation}
{\displaystyle h_{1}j_{5}-\lambda\beta=0},\label{eq:nov23.3}
\end{equation}
where
\[
j_{5}=\frac{\lambda_{\eta}}{\psi_{\eta}}+\lambda^{2}.
\]
As $\lambda=\lambda(\psi)$, then $j_{5}=j_{5}(\psi)$.

Consider $j_{5}\neq0$. Substituting ${\displaystyle h_{1}=\beta\frac{\lambda}{j_{5}}}$
into (\ref{eq:oct12.5}), one gets
\[
\beta(\lambda j_{5\eta}+2j_{5}(\lambda^{2}-j_{5})\psi_{\eta})=0.
\]

If $\lambda j_{5\eta}+2j_{5}(\lambda^{2}-j_{5})\psi_{\eta}\neq0$,
then $\beta=0$ or
\[
k_{5}=\tilde{k}_{2}\frac{k(2\gamma-1)+1}{2k(\gamma-1)^{2}}.
\]
The extension of the kernel of admitted Lie algebras is defined by
the generator
\begin{equation}
\begin{array}{c}
X_{9}^{(5)}=\frac{4S(\gamma-1)^{2}}{j_{1}}\left(-\frac{\partial}{\partial\xi}+g\frac{\partial}{\partial\eta}\right)+(k(2\gamma-1)+1)t\frac{\partial}{\partial t}\\
+(k\gamma+2-\gamma)\left(\varphi\frac{\partial}{\partial\varphi}+\zeta\frac{\partial}{\partial\zeta}+\chi\frac{\partial}{\partial\chi}\right).
\end{array}\label{eq:nov23.1}
\end{equation}

If $\lambda j_{5\eta}+2j_{5}(\lambda^{2}-j_{5})\psi_{\eta}=0$, then
the extension of the kernel of admitted Lie algebras is defined by
the generator $X_{9}^{(6)}=X_{9}^{(5)}$ and one more generator
\begin{equation}
\begin{array}{c}
X_{10}^{(6)}=\frac{2(\gamma-1)\lambda}{(\gamma-2)\psi_{\eta}j_{1}j_{5}}\left((S_{\eta}+2(\gamma-1)\lambda S\psi_{\eta})\frac{\partial}{\partial\xi}-(S_{\xi}+2(\gamma-1)\lambda S\psi_{\xi})\frac{\partial}{\partial\eta}\right)\\
-\frac{1}{\gamma-2}t\frac{\partial}{\partial t}+\varphi\frac{\partial}{\partial\varphi}+\zeta\frac{\partial}{\partial\zeta}+\chi\frac{\partial}{\partial\chi}.
\end{array}\label{eq:nov23.1-1}
\end{equation}


Considering $j_{5}=0$, one obtains that $\lambda\beta=0$.

If $\lambda\neq0$, then ${\displaystyle \beta=0}$ and the extension
of the kernel of admitted Lie algebras is defined by the generator
$X_{9}^{(7)}=X_{9}^{(5)}$ and by one more generator
\begin{equation}
X_{10}^{(7)}=\frac{h_{11}}{\psi_{\eta}j_{1}}\left(-(S_{\eta}+2(\gamma-1)\lambda S\psi_{\eta})\frac{\partial}{\partial\xi}+(S_{\xi}+2(\gamma-1)\lambda S\psi_{\xi})\frac{\partial}{\partial\eta}\right),\label{eq:nov23.5}
\end{equation}
where $h_{11}(\psi)$ is the general solution of equation (\ref{eq:oct12.5}):

\begin{equation}
{\displaystyle h_{11}^{\prime}=h_{11}\lambda.}\label{eq:nov23.6}
\end{equation}

If $\lambda=0$, then solving equation (\ref{eq:oct12.5}), one derives
\begin{equation}
{\displaystyle h_{1}=-\beta\psi+k_{20}},\label{eq:nov23.7}
\end{equation}
where $k_{20}$ is an arbitrary constant. The extension of the kernel
of admitted Lie algebras is defined by the generators
\begin{equation}
\begin{array}{c}
X_{9}^{(8)}={\displaystyle \frac{1}{(\gamma-1)k\psi_{\eta}j_{1}}\left(-((k(2\gamma-1)+1)S_{\eta}\psi+2(\gamma-1)S\psi_{\eta})\frac{\partial}{\partial\xi}\right.}\\
{\displaystyle \left.+((k(2\gamma-1)+1)S_{\xi}\psi+2(\gamma-1)S\psi_{\xi})\frac{\partial}{\partial\eta}\right)}\\
{\displaystyle +\varphi\frac{\partial}{\partial\varphi}+\zeta\frac{\partial}{\partial\zeta}+\chi\frac{\partial}{\partial\chi},}
\end{array}\label{eq:nov23.10}
\end{equation}
\begin{equation}
X_{10}^{(8)}=\frac{2(\gamma-1)\psi}{(\gamma-2)\psi_{\eta}j_{1}}\left(-S_{\eta}\frac{\partial}{\partial\xi}+S_{\xi}\frac{\partial}{\partial\eta}\right),\label{eq:nov23.5-1}
\end{equation}

\[
X_{11}^{(8)}=\frac{1}{\psi_{\eta}j_{1}}\left(-S_{\eta}\frac{\partial}{\partial\xi}+S_{\xi}\frac{\partial}{\partial\eta}\right),
\]


\subsubsection{Case $j_{1}\protect\neq0$ and $j_{2}=0$.}

Equation (\ref{eq:oct8.3}) becomes
\begin{equation}
{\displaystyle h_{1}N-\frac{1}{\gamma-1}(k_{2}+k_{5}(2\gamma-3))=0,}\label{eq:oct12.11}
\end{equation}
where
\[
N=\frac{1}{\psi_{\eta}}\left(\frac{(2\gamma-1)S_{\eta}}{2(\gamma-1)S}-\frac{j_{1\eta}}{j_{1}}-\frac{\psi_{\eta\eta}}{\psi_{\eta}}\right).
\]
Conditions (\ref{eq:oct14.1}) provide that $N=N(\psi)$.


Assume that $N=0$. Finding $\psi_{\eta\eta}$ from the condition
$N=0$:
\begin{equation}
\psi_{\eta\eta}=\psi_{\eta}\left(\frac{(2\gamma-1)S_{\eta}}{2(\gamma-1)S}-\frac{j_{1\eta}}{j_{1}}\right),\label{eq:oct14.2}
\end{equation}
one checks that $(\psi_{\xi})_{\eta\eta}=(\psi_{\eta\eta})_{\xi}$.
Equation (\ref{eq:oct12.11}) reduces to the equation

\[
k_{2}=k_{5}\left(\frac{3}{2}-\gamma\right),
\]
and equation (\ref{eq:oct15.1}) becomes
\[
\left(\frac{h_{1\eta}}{\psi_{\eta}}\right)_{\eta}=0.
\]
\[
k_{\eta}=k_{1}\psi_{\eta}
\]
As $h_{1}=h_{1}(\psi)$, one finds that
\[
h_{1}=k_{21}\psi+k_{20,}
\]
where $k_{21}$ and $k_{20}$ are arbitrary constants. The extension
of the kernel of admitted Lie algebras (\ref{eq:kernel}) is defined
by the generators

\begin{equation}
X_{9}^{(9)}=\frac{4(\gamma-1)^{2}S}{(\gamma-2)j_{1}}\left(\frac{\partial}{\partial\xi}-g\frac{\partial}{\partial\eta}\right)+\varphi\frac{\partial}{\partial\varphi}+\zeta\frac{\partial}{\partial\zeta}+\chi\frac{\partial}{\partial\chi}-\frac{t}{(\gamma-2)}\frac{\partial}{\partial t},\label{eq:nov23.10-3}
\end{equation}

\begin{equation}
X_{10}^{(9)}=\frac{1}{\psi_{\eta}j_{1}}\left(-(\psi S_{\eta}+2(\gamma-1)S\psi_{\eta})\frac{\partial}{\partial\xi}+(\psi S_{\xi}+2(\gamma-1)S\psi_{\xi})\frac{\partial}{\partial\eta}\right),\label{eq:nov23.10-1}
\end{equation}

\begin{equation}
X_{11}^{(9)}=\frac{1}{\psi_{\eta}j_{1}}\left(-S_{\eta}\frac{\partial}{\partial\xi}+S_{\xi}\frac{\partial}{\partial\eta}\right)\label{eq:nov23.10-2}
\end{equation}


Assuming that $N\neq0$, one can introduce the function $P(\psi)$
instead of the function $N(\psi)$ by the formula

\[
P=\frac{1}{(\gamma-1)N}
\]
or the function $P$ is introduced by the formula
\begin{equation}
\psi_{\eta\eta}=\psi_{\eta}\left(\frac{(2\gamma-1)S_{\eta}}{2(\gamma-1)S}-\frac{j_{1\eta}}{j_{1}}-\frac{\psi_{\eta}}{(\gamma-1)P}\right).\label{eq:oct14.3}
\end{equation}
As in the previous case the compatibility condition $(\psi_{\xi})_{\eta\eta}=(\psi_{\eta\eta})_{\xi}$
is also satisfied. Equation (\ref{eq:oct12.11}) gives

\[
h_{1}=\frac{2k_{2}+k_{5}(2\gamma-3)}{2}P.
\]
As $P=P(\psi)$, then $\left(\frac{P_{\eta}}{\psi_{\eta}}\right)_{\eta}=P^{\prime\prime}\psi_{\eta}$,
and equation (\ref{eq:oct15.1}) reduces to
\[
P^{\prime\prime}(2k_{2}+k_{5}(2\gamma-3))=0.
\]

If $P^{\prime\prime}\neq0$, then the extension of the kernel of admitted
Lie algebras (\ref{eq:kernel}) is defined by the generator $X_{9}^{(9)}$

\begin{equation}
X_{9}^{(10)}=\frac{4(\gamma-1)^{2}S}{(\gamma-2)j_{1}}\left(\frac{\partial}{\partial\xi}-g\frac{\partial}{\partial\eta}\right)+\varphi\frac{\partial}{\partial\varphi}+\zeta\frac{\partial}{\partial\zeta}+\chi\frac{\partial}{\partial\chi}-\frac{t}{(\gamma-2)}\frac{\partial}{\partial t},\label{eq:nov23.10-3-1}
\end{equation}
and if $P^{\prime\prime}=0$, then there is one more admitted generator
\begin{equation}
\begin{array}{c}
X_{10}^{(10)}={\displaystyle \frac{1}{\psi_{\eta}j_{1}}\left(-P\left(S_{\eta}\frac{\partial}{\partial\xi}-S_{\xi}\frac{\partial}{\partial\eta}\right)+2S((\gamma-1)P^{\prime}-(2\gamma-1)))\left(\psi_{\eta}\frac{\partial}{\partial\xi}-\psi_{\xi}\frac{\partial}{\partial\eta}\right)\right)}\\
{\displaystyle +\varphi\frac{\partial}{\partial\varphi}+\zeta\frac{\partial}{\partial\zeta}+\chi\frac{\partial}{\partial\chi}.}
\end{array}\label{eq:nov23.10-4}
\end{equation}


\subsection{Case $j_{1}=0$}

The condition $j_{1}=0$ provides that $S=S(\psi)$, equation (\ref{eq:oct15.2})
is satisfied, and equation (\ref{eq:oct15.1}) becomes
\begin{equation}
(\xi^{\eta}+\xi^{\xi}g)S_{\eta}g_{1}+S(2k_{2}(1-2\gamma)+k_{5})=0,\label{eq:oct16.1}
\end{equation}
where
\[
g_{1}=2(\gamma-1)S\frac{\psi_{\eta\eta}S_{\eta}-S_{\eta\eta}\psi_{\eta}}{\psi_{\eta}^{3}}+(2\gamma-1)\frac{S_{\eta}^{2}}{\psi_{\eta}^{2}}
\]

Assume that $g_{1}\neq0$. From the latter equation one finds
\[
\xi^{\eta}=-\xi^{\xi}g-\frac{S}{g_{1}S_{\eta}}(2(2\gamma-1)k_{2}-k_{5}).
\]
Substituting $\xi^{\eta}$ into (\ref{eq:oct15.3}) and (\ref{eq:oct15.4}),
they reduce to the single equation
\[
g_{1\eta}(2(2\gamma-1)k_{2}-k_{5})=0.
\]

Let $g_{1\eta}\neq0$, then $k_{5}=2(2\gamma-1)k_{2}$, and equation
(\ref{eq:oct15.5}) reduces to the quasilinear first-order partial
differential equation for the single function $\xi^{\xi}$:
\[
\xi_{\xi}^{\xi}-g\xi_{\eta}^{\xi}=g_{\eta}\xi^{\xi}+4(\gamma-1)k_{2}.
\]
The general solution of the latter equation can be written as follows
\[
\xi^{\xi}=\psi_{\eta}\left(h_{11}-4(\gamma-1)k_{2}h_{12}\right),
\]
where $h_{11}=h_{11}(\psi)$ is an arbitrary function and $h_{12}(\xi,\eta)$
is an arbitrary solution of the linear equation
\[
h_{12\xi}-gh_{12\eta}+\psi_{\eta}^{-1}=0.
\]
The extension of the kernel of admitted Lie algebras (\ref{eq:kernel})
is defined by the generators

\begin{equation}
X_{9}^{(12)}=h_{11}\psi_{\eta}\left(\frac{\partial}{\partial\xi}-g\frac{\partial}{\partial\eta}\right),\label{eq:oct16.11}
\end{equation}

\begin{equation}
X_{10}^{(12)}=\begin{array}{c}
{\displaystyle \frac{2(\gamma-1)}{\gamma}h_{12}\left(-\psi_{\eta}\frac{\partial}{\partial\xi}+\psi_{\xi}\frac{\partial}{\partial\eta}\right)+\varphi\frac{\partial}{\partial\varphi}+\zeta\frac{\partial}{\partial\zeta}+\chi\frac{\partial}{\partial\chi}+\frac{2\gamma-1}{\gamma}t\frac{\partial}{\partial t}.}\end{array}\label{eq:oct16.12}
\end{equation}


Let $g_{1\eta}=0$, say $g_{1}=k$, where $k\neq0$ is constant. Equation
(\ref{eq:oct15.5}) becomes
\[
\begin{array}{c}
\xi_{\xi}^{\xi}-g\xi_{\eta}^{\xi}=g_{\eta}\xi^{\xi}+\frac{S\psi_{\eta\eta}}{kS_{\eta}\psi_{\eta}}\left(2k_{2}(2\gamma-1)-k_{5}\right)\\
+\frac{1}{2k(\gamma-1)}(2k_{2}(k+1-2\gamma)+k_{5}(k(2\gamma-3)+1).
\end{array}
\]

The general solution of the latter equation is written in the form
\[
\xi^{\xi}=\psi_{\eta}(h_{11}+k_{2}h_{12}+k_{5}h_{13}),
\]
where $h_{11}=h_{11}(\psi)$ is an arbitrary function, $h_{12}(\xi,\eta)$
and $h_{13}(\xi,\eta)$ are arbitrary solutions of the linear equations
\[
h_{12\xi}-gh_{12\eta}=\frac{(1-\gamma)(2\gamma-1)\psi_{\eta\eta}S}{kS_{\eta}\psi_{\eta}^{2}}+\frac{2\gamma-1-k}{2k\psi_{\eta}},
\]
\[
h_{13\xi}-gh_{13\eta}=\frac{(1-\gamma)\psi_{\eta\eta}S}{kS_{\eta}\psi_{\eta}^{2}}+\frac{k(2\gamma-3)+1}{2k\psi_{\eta}}.
\]
The extension of the kernel of admitted Lie algebras (\ref{eq:kernel})
is defined by the generators $X_{9}^{(13)}=X_{9}^{(12)}$ and

\[
X_{10}^{(13)}=\begin{array}{c}
\frac{2h_{12}}{\gamma-1}\left(-\psi_{\eta}\frac{\partial}{\partial\xi}+\psi_{\xi}\frac{\partial}{\partial\eta}\right)+\frac{2(2\gamma-1)S}{kS_{\eta}}\frac{\partial}{\partial\eta}+\varphi\frac{\partial}{\partial\varphi}+\zeta\frac{\partial}{\partial\zeta}+\chi\frac{\partial}{\partial\chi}.\end{array}
\]

\[
X_{11}^{(13)}=\begin{array}{c}
\frac{2h_{13}}{\gamma-1}\left(\psi_{\eta}\frac{\partial}{\partial\xi}-\psi_{\xi}\frac{\partial}{\partial\eta}\right)-\frac{2S}{kS_{\eta}}\frac{\partial}{\partial\eta}+\varphi\frac{\partial}{\partial\varphi}+\zeta\frac{\partial}{\partial\zeta}+\chi\frac{\partial}{\partial\chi}+2t\frac{\partial}{\partial t}.\end{array}
\]


Case $g_{1}=0$. Equation (\ref{eq:oct16.1}) gives that $k_{5}=2k_{2}(2\gamma-1)$,
and equation (\ref{eq:oct15.5}) takes the form
\begin{equation}
\xi_{\xi}^{\xi}-g\xi_{\eta}^{\xi}=(g_{\eta}+g\frac{g_{2\eta}}{g_{2}})\xi^{\xi}+\frac{g_{2\eta}}{g_{2}}\xi^{\eta}+4k_{2}(\gamma-1),\label{eq:oct16.3}
\end{equation}
where
\begin{equation}
\frac{g_{2\eta}}{g_{2}}=\frac{\psi_{\eta\eta}}{\psi_{\eta}}-\frac{S_{\eta}}{2(\gamma-1)S},\label{eq:oct16.15}
\end{equation}
where such denotation is introduced for further simplifications.

Case $g_{2\eta}\neq0$. Finding $\xi^{\eta}$ from equation (\ref{eq:oct16.3}),
and substituting it into equations (\ref{eq:oct15.3}) and (\ref{eq:oct15.4}),
one obtains two second-order equations for $\xi^{\xi}$. These equations
can be simplified by the substitution
\[
\xi_{\xi}^{\xi}=g\xi_{\eta}^{\xi}+g_{\eta}\xi^{\xi}+4(\gamma-1)k_{2}+\frac{g_{2\eta}}{g_{2}^{2}}S^{-1/(\gamma-1)}\left(k_{2}+h\right),
\]
where $h(\xi,\eta)$ is some unknown function. Equations (\ref{eq:oct15.3})
and (\ref{eq:oct15.4}) become, respectively,
\begin{equation}
h_{\xi}=0,\,\,\,h_{\eta}=0.\label{eq:oct16.14}
\end{equation}
Hence, $h$ is constant, say $h=k_{20}$.

The general solution of the latter equation is presented in the form
\[
\xi^{\xi}=\psi_{\eta}(h_{11}+4(\gamma-1)k_{2}h_{12}+h_{13}k_{20}),
\]
where $h_{11}=h_{11}(\psi)$ is arbitrary function, and $h_{12}(\xi,\eta)$
and $h_{13}(\xi,\eta)$ are arbitrary solutions of the linear equations
\begin{equation}
\begin{array}{c}
h_{12\xi}-gh_{12\eta}=\psi_{\eta}^{-1}\left(\frac{g_{2\eta}}{4(\gamma-1)g_{2}^{2}}S^{-1/(\gamma-1)}+1\right).\\
h_{13\xi}-gh_{12\eta}=S^{-1/(\gamma-1)}\frac{g_{2\eta}}{\psi_{\eta}g_{2}^{2}}.
\end{array}\label{eq:oct16.4-1}
\end{equation}
The extension of the kernel of admitted Lie algebras (\ref{eq:kernel})
is defined by the generators (\ref{eq:oct16.11}) and (\ref{eq:oct16.12}):

\[
X_{9}^{(14)}=h_{11}\left(\psi_{\eta}\frac{\partial}{\partial\xi}-\psi_{\xi}\frac{\partial}{\partial\eta}\right),
\]

\[
X_{10}^{(14)}=\begin{array}{c}
{\displaystyle \frac{2(\gamma-1)}{\gamma}h_{12}\left(\psi_{\eta}\frac{\partial}{\partial\xi}-\psi_{\xi}\frac{\partial}{\partial\eta}\right)+\frac{1}{2\gamma g_{2}}S^{-1/(\gamma-1)}\frac{\partial}{\partial\eta}+\varphi\frac{\partial}{\partial\varphi}+\zeta\frac{\partial}{\partial\zeta}+\chi\frac{\partial}{\partial\chi}+\frac{2\gamma-1}{\gamma}t\frac{\partial}{\partial t},}\end{array}
\]
\[
X_{11}^{(14)}=h_{13}\left(\psi_{\eta}\frac{\partial}{\partial\xi}-\psi_{\xi}\frac{\partial}{\partial\eta}\right)+\frac{1}{g_{2}}S^{-1/(\gamma-1)}\frac{\partial}{\partial\eta}.
\]

Let $g_{2\eta}=0$.
\[
\frac{\psi_{\eta\eta}}{\psi_{\eta}}-\frac{S_{\eta}}{2(\gamma-1)S}=0\Rightarrow\left(\psi_{\eta}S^{-1/(2(\gamma-1))}\right)_{\eta}=0.
\]
The compatibility condition $(\psi_{\eta\eta})_{\xi}=(\psi_{\xi})_{\eta\eta}$
gives that $g(\xi,\eta)$ is a linear function with respect to $\eta$,
say
\[
g=-\frac{\mu_{1}^{\prime\prime}}{\mu_{1}^{\prime}}\eta+\mu_{2}^{\prime}\mu_{1}^{\prime},
\]
where $\mu_{1}(\xi)$ and $\mu_{2}(\xi)$ are some functions such
that $\mu_{1}^{\prime}\neq0$. Here the representation for $g$ is
chosen for convenience of further integration. In particular, solving
the equation $\psi_{\xi}=g\psi_{\eta}$, one finds
\[
\psi=\psi(z),\,\,\,z=\frac{\eta}{\mu_{1}^{\prime}}+\mu_{2}.
\]
The relation $g_{2}=0$ provides that
\[
\psi^{\prime}=qS^{\frac{1}{2(\gamma-1)}},
\]
where $q$ is constant. Notice that if one considers $S=S(z)$, then
the relation $g_{2\eta}=0$ gives
\[
S^{\prime}=\frac{2(\gamma-1)S\psi^{\prime\prime}}{\psi^{\prime}\,^{2}}.
\]
The condition $g_{1}=0$ leads to the relation
\[
\psi^{\prime}\psi^{\prime\prime\prime}-3\psi^{\prime\prime}{}^{2}=0.
\]

Introducing $h_{1}=\xi^{\eta}+g\xi^{\xi}$, one derives
\[
\xi^{\eta}=h_{1}-g\xi^{\xi}.
\]
Then equation (\ref{eq:oct15.3}) reduces to
\[
\left(h_{1}S^{1/(\gamma-1)}\right)_{\eta}=0,
\]
which gives
\[
h_{1}=\mu_{3}S^{-1/(\gamma-1)},
\]
where $\mu_{3}(\xi)$ is an arbitrary function. Substituting $h_{1}$
into equation (\ref{eq:oct15.4}), one obtains that $\mu_{3}=k_{20}\mu_{1}^{\prime}$
with constant $k_{20}$. Equation (\ref{eq:oct15.5}) takes the form
\[
\xi_{\xi}^{\xi}+\left(\frac{\mu_{1}^{\prime\prime}}{\mu_{1}^{\prime}}\eta-\mu_{2}^{\prime}\mu_{1}^{\prime}\right)\xi_{\eta}^{\xi}=-\frac{\mu_{1}^{\prime\prime}}{\mu_{1}^{\prime}}\xi^{\xi}+4(\gamma-1)k_{2}.
\]
The general solution of the latter equation is
\[
\xi^{\xi}=k_{2}\frac{4(\gamma-1)\mu_{1}}{\mu_{1}^{\prime}}+\frac{1}{\mu_{1}^{\prime}}F(z),
\]
where $F(z)$ is an arbitrary function.

The extension of the kernel of admitted Lie algebras (\ref{eq:kernel})
is defined by the generators

\[
X_{9}^{(16)}=F\left(z_{\eta}\frac{\partial}{\partial\xi}-z_{\xi}\frac{\partial}{\partial\eta}\right),
\]

\[
{\displaystyle \begin{array}{c}
{\displaystyle X_{10}^{(16)}=\frac{2(\gamma-1)}{\gamma}\mu_{1}\left(z_{\eta}\frac{\partial}{\partial\xi}-z_{\xi}\frac{\partial}{\partial\eta}\right)+\varphi\frac{\partial}{\partial\varphi}+\zeta\frac{\partial}{\partial\zeta}+\chi\frac{\partial}{\partial\chi}+\frac{2\gamma-1}{\gamma}t\frac{\partial}{\partial t},}\end{array}}
\]

\[
X_{11}^{(16)}=\mu_{1}^{\prime}S^{-1/(\gamma-1)}\frac{\partial}{\partial\eta},
\]

\section{Nonisentropic case with $b_{01}^{2}+b_{02}^{2}=0$}

\label{nonisentr_b01_eq0}

In this case $H_{1}=0$ and $H_{2}=0$, and equation (\ref{eq:3_direction})
is integrated
\[
\chi=t\chi_{1}+\chi_{0},
\]
where $\chi_{0}(\xi,\eta)$ and $\chi_{1}(\xi,\eta)$ are arbitrary
functions. Then the variable $\chi(t,\xi,\eta)$ is excluded from
the consideration. It is also assumed that $S_{\eta}\neq0$. Partially
solving the determining equations, one derives that $\xi^{\xi}=\xi^{\xi}(\xi,\eta)$,
$\xi^{\eta}=\xi^{\eta}(\xi,\eta)$, and
\[
\zeta^{\varphi}=k_{1}\zeta+(k_{2}+\frac{1}{2}k_{5})\varphi+tk_{7}+k_{8},
\]
\[
\zeta^{\zeta}=(k_{2}+\frac{1}{2}k_{5})\zeta-k_{1}\varphi+tk_{9}+k_{10},
\]
\[
\zeta^{\chi}=\chi\zeta^{1}+t\zeta^{2}+\zeta^{3},
\]
\[
\xi^{t}=k_{5}t+k_{6},
\]
where $\zeta^{i}(\xi,\eta),\,\,\,(i=1,2,3,4)$ are arbitrary functions.
The kernel of admitted Lie algebras (\ref{eq:kernel}) in these cases
is extended by the generators of the form
\[
Y=(\chi\zeta^{1}+t\zeta^{2}+\zeta^{3})\frac{\partial}{\partial\chi}.
\]
The remaining equations are

\begin{equation}
S(\xi^{\eta}S_{\eta}+\xi^{\xi}S_{\xi})_{\xi}-S_{\xi}(\xi^{\eta}S_{\eta}+\xi^{\xi}S_{\xi})=0,\label{eq:nov11.14}
\end{equation}
\begin{equation}
\begin{array}{c}
\begin{array}{c}
{\displaystyle S(\xi_{\eta}^{\xi}S_{\xi}-\xi_{\xi}^{\xi}S_{\eta})+\xi^{\xi}\left(SS_{\xi\eta}-\frac{\gamma}{\gamma-1}S_{\xi}S_{\eta}\right)+\xi^{\eta}\left(SS_{\eta\eta}-\frac{\gamma}{\gamma-1}S_{\eta}^{2}\right)}\\
+SS_{\eta}\left(2k_{2}+\frac{\gamma}{\gamma-1}k_{5}\right)=0,
\end{array}\end{array}\label{eq:nov11.11}
\end{equation}

\begin{equation}
\begin{array}{c}
\xi^{\eta}\left((\gamma-1)b_{03\eta}-\frac{S_{\eta}}{2S}b_{03}\right)+\xi^{\xi}\left((\gamma-1)b_{03\xi}-\frac{S_{\xi}}{2S}b_{03}\right)\\
-b_{03}(k_{2}(\gamma-2)+k_{5}(1-\frac{\gamma}{2}))=0,
\end{array}\label{eq:nov11.13}
\end{equation}

\begin{equation}
\xi_{\eta}^{\eta}+\xi_{\xi}^{\xi}+\frac{1}{(\gamma-1)S}(S_{\eta}\xi^{\eta}+S_{\xi}\xi^{\xi})=\frac{1}{\gamma-1}(2\gamma k_{2}+(\gamma-2)k_{5}).\label{eq:nov11.15}
\end{equation}
Notice that if $S$ is not constant, then the latter equation can
be reduced to the form
\[
S(\xi^{\eta}S_{\eta}+\xi^{\xi}S_{\xi})_{\eta}-S_{\eta}(\xi^{\eta}S_{\eta}+\xi^{\xi}S_{\xi})=0.
\]

Let $h_{\ensuremath{1}}=\xi^{\xi}S_{\xi}+\xi^{\eta}S_{\eta}$, $h_{2}=\xi^{\xi}b_{03\xi}+\xi^{\eta}b_{03\eta}$,
and $f_{2}=b_{03\xi}S_{\eta}-b_{03\eta}S_{\xi}.$

\subsection{Case $f_{2}\protect\neq0$}


One can derive
\[
\xi^{\xi}=f_{2}^{-1}(h_{2}S_{\eta}-h_{1}b_{03\eta}),\,\,\,\xi^{\eta}=f_{2}^{-1}(-h_{2}S_{\xi}+h_{1}b_{03\xi}),\,\,
\]
From equations (\ref{eq:nov11.14}) and (\ref{eq:nov11.15}) one finds
\[
h_{1}=Sk_{20},
\]
where $k_{20}$ is constant. Equation (\ref{eq:nov11.13}) gives
\[
h_{2}=b_{03}\frac{(\gamma-2)(2k_{2}-k_{5})+k_{20}}{2(\gamma-1)}.
\]

Let $k_{2}=\frac{(\gamma-2)k_{5}+\tilde{k}_{2}}{2(\gamma-2)}$. Equation
(\ref{eq:nov11.11}) becomes
\begin{equation}
k_{2}=f_{3}\tilde{k}_{2}+bk_{20}\label{eq:oct28.1}
\end{equation}
where
\[
f_{3}=\frac{1}{4(\gamma-1)}\left(\frac{b_{03}}{f_{2}^{2}}(S_{\xi}f_{2\eta}-S_{\eta}f_{2\xi})-\frac{\gamma+2}{\gamma-2}\right),\,\,\,b=f_{4}+f_{3}+\frac{\gamma}{2(\gamma-2)},
\]
and
\[
f_{4}=\frac{S}{S_{\eta}}\left(\frac{b_{03\eta}}{b_{03}}\left(2(1-\gamma)f_{3}-\frac{\gamma+2}{2(\gamma-2)}\right)-\frac{f_{2\eta}}{2f_{2}}\right).
\]

Differentiating $k_{2}$ with respect to $\xi$ and $\eta$, one derives
that it is necessary to study the cases (a) $f_{3}\neq const$ and
(b) $f_{3}=const$.


If $f_{3}\neq const$, then one can assume that $f_{3}\neq0$. Hence,
for existence of an extension of the kernel of admitted Lie algebras
one obtains from equation (\ref{eq:oct28.1}) that there exist constants
$k$ and $q$ such that $b=kf_{3}+q$ and $\tilde{k}_{2}=-kk_{20}$.

Thus, the extension of the kernel of admitted Lie algebras is defined
by the generator
\[
\begin{array}{c}
X_{9}^{(17)}={\displaystyle \left(q-\frac{k}{2(\gamma-2)}\right)(\varphi\frac{\partial}{\partial\varphi}+\zeta\frac{\partial}{\partial\zeta})+qt\frac{\partial}{\partial t}}\\
{\displaystyle +\frac{1}{2f_{2}}\left(-\left(2Sb_{03\eta}+(k-1)b_{03}S_{\eta}\right)\frac{\partial}{\partial\xi}+\left(2Sb_{03\xi}+(k-1)b_{03}S_{\xi}\right)\frac{\partial}{\partial\eta}\right).}
\end{array}
\]


Notice that as $S_{\eta}\neq0$, then from the definition of $f_{3}$
one can find $f_{2\xi}$. Finding $f_{2\eta}$ from the equation $b=kf_{3}+q$,
the compatibility condition $(f_{2\xi})_{\eta}-(f_{2\eta})_{\xi}=0$
gives
\[
2(\gamma-1)S(b_{03\eta}f_{3\xi}-b_{03\xi}f_{3\eta})+(k-1)b_{03}(S_{\eta}f_{3\xi}-S_{\xi}f_{3\eta})=0.
\]




Consider case $f_{3}=const$.

Assume that $f_{4}$ is constant, say $f_{4}=m$. In this case $f_{2}=qb_{03}^{q_{2}}S^{q_{3}}$,
where $q$ is an arbitrary constant and
\[
q_{2}=-4(\gamma-1)f_{3}-\frac{\gamma+2}{\gamma-2},\,\,\,q_{3}=-2m.
\]

Thus,
\[
b_{03\xi}=(b_{03\eta}S_{\xi}+f_{2})/S_{\eta},
\]
and the extension of admitted Lie algebras occurs by the generators
\[
\begin{array}{c}
X_{9}^{(18)}={\displaystyle \left(f_{3}+m+\frac{\gamma}{2(\gamma-2)}\right)\left(\varphi\frac{\partial}{\partial\varphi}+\zeta\frac{\partial}{\partial\zeta}+t\frac{\partial}{\partial t}\right)}\\
{\displaystyle +\frac{1}{2qb_{03}^{q_{2}}S^{q_{3}}}\left(\left(-2Sb_{03\eta}+\frac{b_{03}S_{\eta}}{\gamma-1}\right)\frac{\partial}{\partial\xi}-\left(-2Sb_{03\xi}+\frac{b_{03}S_{\xi}}{\gamma-1}\right)\frac{\partial}{\partial\eta}\right),}
\end{array}
\]
\[
\begin{array}{c}
X_{10}^{(18)}={\displaystyle \left(f_{3}+\frac{1}{2(\gamma-2)}\right)\left(\varphi\frac{\partial}{\partial\varphi}+\zeta\frac{\partial}{\partial\zeta}\right)+f_{3}t\frac{\partial}{\partial t}}\\
{\displaystyle +\frac{S^{-q_{3}}b_{03}^{1-q_{2}}}{2q(\gamma-1)}\left(S_{\eta}\frac{\partial}{\partial\xi}-S_{\xi}\frac{\partial}{\partial\eta}\right).}
\end{array}
\]



If $f_{4}$ is not constant, then $k_{20}=0$, $f_{4}=f_{4}(S)$,
and
\[
b_{03\xi}=(b_{03\eta}S_{\xi}+f_{2})/S_{\eta}.
\]
The extension of the kernel of admitted Lie algebras consists of the
generator 
\[
X_{9}^{(19)}=\left(f_{3}+\frac{1}{2(\gamma-2)}\right)\left(\varphi\frac{\partial}{\partial\varphi}+\zeta\frac{\partial}{\partial\zeta}\right)+f_{3}t\frac{\partial}{\partial t}+\frac{b_{03}}{2(\gamma-1)f_{2}}\left(S_{\eta}\frac{\partial}{\partial\xi}-S_{\xi}\frac{\partial}{\partial\eta}\right).
\]


\subsection{Case $f_{2}=0$}

In this case $b_{03}=b_{03}(S)$. From equations (\ref{eq:nov11.14})
and (\ref{eq:nov11.15}) one finds that $h_{1}=Sk_{20}$. Equation
(\ref{eq:nov11.13}) becomes k
\[
k_{5}=2\tilde{k}_{2}+f_{5}k_{20},
\]
where

\[
f_{5}=\frac{2(1-\gamma)}{(\gamma-2)}\frac{Sb_{03\eta}}{S_{\eta}b_{03}}+\frac{1}{\gamma-2}.
\]
Notice that $f_{5}=f_{5}(S)$.


If $f_{5}\neq const$, then $k_{20}=0$, the general solution of equation
(\ref{eq:nov11.11}) can be presented in the form

\[
\xi^{\xi}=S_{\eta}(\psi_{1}+\psi_{2}k_{2}).
\]
Substituting the latter into (\ref{eq:nov11.11}), one finds that
$\psi_{1}(S)$ is an arbitrary function and the function $\psi_{2}$
is a solution of the equation
\[
\psi_{2\xi}S_{\eta}-\psi_{2\eta}S_{\xi}=4.
\]
The extension of the kernel of admitted Lie algebras occurs by the
generators 
\[
X_{9}^{(20)}=\psi_{1}\left(S_{\eta}\frac{\partial}{\partial\xi}-S_{\xi}\frac{\partial}{\partial\eta}\right),
\]
\[
X_{10}^{(20)}=2\left(\varphi\frac{\partial}{\partial\varphi}+\zeta\frac{\partial}{\partial\zeta}+t\frac{\partial}{\partial t}\right)+\psi_{2}\left(S_{\eta}\frac{\partial}{\partial\xi}-S_{\xi}\frac{\partial}{\partial\eta}\right).
\]


If $f_{5}=k$, then the extension of the kernel of admitted Lie algebras
is defined by the generators $X_{9}^{21}=X_{9}^{20}$ and $X_{10}^{21}=X_{10}^{20}$
and by one more generator
\[
X_{11}^{(21)}=k\left(\varphi\frac{\partial}{\partial\varphi}+\zeta\frac{\partial}{\partial\zeta}+2t\frac{\partial}{\partial t}\right)+2\psi_{3}\left(S_{\eta}\frac{\partial}{\partial\xi}-S_{\xi}\frac{\partial}{\partial\eta}\right)+2\frac{S}{S_{\eta}}\frac{\partial}{\partial\eta},
\]
where the function $\psi_{3}$ is a solution of the equation
\[
S_{\eta}\psi_{3\xi}-S_{\xi}\psi_{3\eta}=\frac{SS_{\eta\eta}}{S_{\eta}^{2}}+\frac{k(\gamma-2)-\gamma}{\gamma-1}.
\]

\section{Isentropic case with $b_{01}^{2}+b_{02}^{2}\protect\neq0$}

\label{isentr_b01_neq0} 


For the isentropic case, equations (\ref{eq:oct15.1})-(\ref{eq:oct15.5})
reduce to the following
\begin{equation}
\xi_{\xi}^{\eta}=-\xi_{\eta}^{\xi}g^{2}+\xi^{\eta}(-g_{\eta}-2\psi_{\eta\eta}\psi_{\eta}^{-1}g)+\xi^{\xi}(-2g_{\eta}g-g_{\xi}-2\psi_{\eta\eta}\psi_{\eta}^{-1}g^{2})+g(2k_{2}-k_{5}),\label{eq:2_jan28}
\end{equation}
\begin{equation}
eq_{1}+eq_{3}=\begin{array}{c}
\xi_{\xi}^{\xi}+\xi_{\eta}^{\eta}=2\frac{2\gamma k_{2}+(\gamma-2)k_{5}}{\gamma-1},\end{array}\label{eq:oct15.4-1}
\end{equation}

\begin{equation}
\xi_{\xi}^{\xi}=\xi_{\eta}^{\xi}g+\xi^{\eta}\psi_{\eta\eta}\psi_{\eta}^{-1}+\xi^{\xi}(g_{\eta}+\psi_{\eta\eta}\psi_{\eta}^{-1}g)+\frac{2\gamma k_{5}+2k_{2}-3k_{5}}{\gamma-1},\label{eq:1_jan28}
\end{equation}


Assume that $\psi_{\eta\eta}\neq0$. Substituting $\xi^{\eta}$, found
from equation (\ref{eq:1_jan28}), into (\ref{eq:2_jan28}) and (\ref{eq:oct15.4-1}),
one can integrate them
\begin{equation}
\xi_{\xi}^{\xi}-\xi_{\eta}^{\xi}g-\xi^{\xi}g_{\eta}+\frac{\psi_{\eta\eta}}{\psi_{\eta}^{2}}\left(\frac{((1-2\gamma)k_{2}+\frac{1}{2}k_{5})\psi}{\gamma-1}+k_{20}\right)-\frac{k_{2}+(\gamma-\frac{3}{2})k_{5}}{\gamma-1}=0,\label{eq:4_jan28}
\end{equation}
The latter is a linear equation for the function $\xi^{\xi}$. The
general solution of this equation can be found in the form
\[
\xi^{\xi}=\psi_{\eta}(\psi_{1}+\psi_{2}k_{2}+\psi_{3}k_{20}+\psi_{4}k_{5}),
\]
where $\psi_{i}=\psi_{i}(\xi,\eta),\ (i=1,2,3,4)$ are some functions.
Substituting $\xi^{\xi}$ into (\ref{eq:4_jan28}) and splitting it,
one obtains
\[
\psi_{1\xi}=\psi_{1\eta}g,
\]
\[
\psi_{2\xi}=\psi_{2\eta}g+\frac{1}{\gamma-1}\left(\psi_{\eta\eta}\psi_{\eta}^{-3}\psi(2\gamma-1)+\psi_{\eta}^{-1}\right),
\]
\[
\psi_{3\xi}=-\psi_{\eta\eta}\psi_{\eta}^{-3}+\psi_{3\eta}g,
\]
\[
\psi_{4\xi}=\psi_{4\eta}g+\frac{1}{2(\gamma-1}\left(\psi_{\eta\eta}\psi_{\eta}^{-3}\psi+\psi_{\eta}^{-1}(2\gamma-3)\right).
\]

The extension of admitted Lie algebras is defined by the generators
\[
X_{9}^{(22)}=\varphi\frac{\partial}{\partial\varphi}+\zeta\frac{\partial}{\partial\zeta}+\chi\frac{\partial}{\partial\chi}+2t\frac{\partial}{\partial t}+2\psi_{4}(\psi_{\eta}\frac{\partial}{\partial\xi}-\psi_{\xi}\frac{\partial}{\partial\eta})-\frac{\psi}{(\gamma-1)\psi_{\eta}}\frac{\partial}{\partial\eta}
\]
\[
X_{10}^{(22)}=\psi_{3}(\psi_{\eta}\frac{\partial}{\partial\xi}-\psi_{\xi}\frac{\partial}{\partial\eta})-\psi_{\eta}^{-1}\frac{\partial}{\partial\eta},
\]
\[
X_{11}^{(22)}=\varphi\frac{\partial}{\partial\varphi}+\zeta\frac{\partial}{\partial\zeta}+\chi\frac{\partial}{\partial\chi}+\psi_{2}(\psi_{\eta}\frac{\partial}{\partial\xi}-\psi_{\xi}\frac{\partial}{\partial\eta})+\frac{(2\gamma-1)\psi}{(\gamma-1)\psi_{\eta}}\frac{\partial}{\partial\eta},
\]
\[
X_{12}^{(22)}=\psi_{1}(\psi_{\eta}\frac{\partial}{\partial\xi}-\psi_{\xi}\frac{\partial}{\partial\eta}).
\]


\[
\psi=\frac{\eta}{\mu_{1}^{\prime}}+\mu_{2}
\]
equation (\ref{eq:1_jan28}) can be integrated
\[
\xi^{\xi}=\frac{1}{(\gamma-1)\mu_{1}^{\prime}}((\gamma-1)h+(\gamma-\frac{3}{2})k_{5}\mu_{1}+k_{2}\mu_{1}),
\]
where $h(\psi)$ is an arbitrary function. Integrating equation (\ref{eq:oct15.4-1}),
one finds
\[
\xi^{\eta}+\psi_{\xi}h-\alpha\eta=f\mu_{1}^{\prime},
\]
where
\[
\alpha=\mu_{1}^{\prime\prime}\frac{\mu_{1}(k_{2}+(\gamma-\frac{3}{2})k_{5})}{(\gamma-1)\mu_{1}^{\prime}{}^{2}}+\frac{(2\gamma-1)k_{2}-\frac{1}{2}k_{5}}{\gamma-1},
\]
and the function $f(\xi)$ is a function of the integration. Substituting
$\xi^{\eta}$ into equation (\ref{eq:2_jan28}), one finds the function
$f$:
\[
f=\alpha_{1}k_{2}+\alpha_{2}k_{5}+k_{20},
\]
where
\[
\alpha_{1}=-\frac{\mu_{2}^{\prime}\mu_{1}}{(\gamma-1)\mu_{1}^{\prime}}+\frac{2\gamma-1}{\gamma-1}\mu_{2},
\]
\[
\alpha_{2}=-\frac{(\gamma-3/2)\mu_{2}^{\prime}\mu_{1}}{(\gamma-1)\mu_{1}^{\prime}}-\frac{\mu_{2}}{2(\gamma-1)},
\]
and $k_{20}$ is a constant of integration.

The extension of admitted Lie algebras is defined by the generators
\[
X_{11}^{(23)}=h\left(\psi_{\eta}\frac{\partial}{\partial\xi}-\psi_{\xi}\frac{\partial}{\partial\eta}\right),
\]
\[
X_{9}^{(23)}=\varphi\frac{\partial}{\partial\varphi}+\zeta\frac{\partial}{\partial\zeta}+\chi\frac{\partial}{\partial\chi}+2t\frac{\partial}{\partial t}+\frac{(2\gamma-3)\mu_{1}}{\gamma-1}(\psi_{\eta}\frac{\partial}{\partial\xi}-\psi_{\xi}\frac{\partial}{\partial\eta})-\frac{\psi\mu_{1}^{\prime}}{\gamma-1}\frac{\partial}{\partial\eta},
\]

\[
X_{10}^{(23)}=\varphi\frac{\partial}{\partial\varphi}+\zeta\frac{\partial}{\partial\zeta}+\chi\frac{\partial}{\partial\chi}+\frac{\mu_{1}}{\gamma-1}(\psi_{\eta}\frac{\partial}{\partial\xi}-\psi_{\xi}\frac{\partial}{\partial\eta})+\frac{(2\gamma-1)\psi\mu_{1}^{\prime}}{\gamma-1}\frac{\partial}{\partial\eta}.
\]


\section{Isentropic case with $b_{01}^{2}+b_{02}^{2}=0$}

\label{isentr_b01_eq0} 


First of all it should be notice that $\gamma=2$ is equivalent to
the gas dynamics equations. Hence, it is assumed that $\gamma\neq2$.
The defining equations (\ref{eq:nov11.14})-(\ref{eq:nov11.15}) reduce
to the equations
\begin{equation}
b_{03\eta}\xi^{\eta}+b_{03\xi}\xi^{\xi}=\frac{b_{03}}{\gamma-1}\left(k_{2}(\gamma-2)-\frac{\gamma-2}{2}k_{5}\right),\label{eq:nov11.13-1}
\end{equation}
\begin{equation}
\xi_{\eta}^{\eta}+\xi_{\xi}^{\xi}=\frac{1}{\gamma-1}(2\gamma k_{2}+(\gamma-2)k_{5}).\label{eq:jan29.1}
\end{equation}


Assume that $b_{03}$ is not constant, for example, $b_{03\eta}\neq0{\normalcolor {\normalcolor }}$.
Finding $\xi^{\eta}$ from (\ref{eq:nov11.13-1}) and substituting
it into (\ref{eq:jan29.1}), one obtains a linear first-order partial
differential equation for the function $\xi^{\xi}$. Representing
the general solution of this equation in the form

\[
\xi^{\xi}=b_{03\eta}(\psi_{1}+\psi_{2}k_{2}+\psi_{3}k_{5}),
\]
one derives
\[
\psi_{1\xi}b_{03\eta}-\psi_{1\eta}b_{03\xi}=0,
\]
\[
\psi_{2\xi}b_{03\eta}-\psi_{2\eta}b_{03\xi}=\frac{(\gamma-2)b_{03}b_{03\eta\eta}}{(\gamma-1)b_{03\eta}^{2}}+\frac{(\gamma+2)}{(\gamma-1)},
\]
\[
\psi_{3\xi}b_{03\eta}-\psi_{3\eta}b_{03\xi}=\frac{(\gamma-2)}{2(\gamma-1)}\left(3-\frac{b_{03}b_{03\eta\eta}}{b_{03\eta}^{2}}\right).
\]
The extension of admitted Lie algebras occurs by the generators 
\[
X_{9}^{(24)}=\psi_{1}\left(b_{03\eta}\frac{\partial}{\partial\xi}-b_{03\xi}\frac{\partial}{\partial\eta}\right),
\]
\[
X_{10}^{(24)}=\varphi\frac{\partial}{\partial\varphi}+\zeta\frac{\partial}{\partial\zeta}+\psi_{2}\left(b_{03\eta}\frac{\partial}{\partial\xi}-b_{03\xi}\frac{\partial}{\partial\eta}\right)+\frac{(\gamma-2)b_{03}}{(\gamma-1)b_{03\eta}}\frac{\partial}{\partial\eta},
\]

\[
X_{11}^{(24)}=2\psi_{3}\left(b_{03\eta}\frac{\partial}{\partial\xi}-b_{03\xi}\frac{\partial}{\partial\eta}\right)-\frac{(\gamma-2)b_{03}}{(\gamma-1)b_{03\eta}}\frac{\partial}{\partial\eta}+\varphi\frac{\partial}{\partial\varphi}+\zeta\frac{\partial}{\partial\zeta}+2t\frac{\partial}{\partial t}.
\]



If $b_{03}$ is constant, then $k_{5}=2k_{2}$, $\xi^{\eta}=\psi_{1\xi}$,
$\xi^{\xi}=-\psi_{1\eta}+4k_{2}\xi$, where $\psi_{1}(\xi,\eta)$
is an arbitrary function, and the extension of admitted Lie algebras
occurs by the generators 
\[
X_{9}^{(25)}=\psi_{1\eta}\frac{\partial}{\partial\xi}-\psi_{1\xi}\frac{\partial}{\partial\eta},\,\,\,X_{10}^{(25)}=\varphi\frac{\partial}{\partial\varphi}+\zeta\frac{\partial}{\partial\zeta}+t\frac{\partial}{\partial t}+2\xi\frac{\partial}{\partial\xi}.
\]


\section*{Conclusions}

\label{conclusion} 
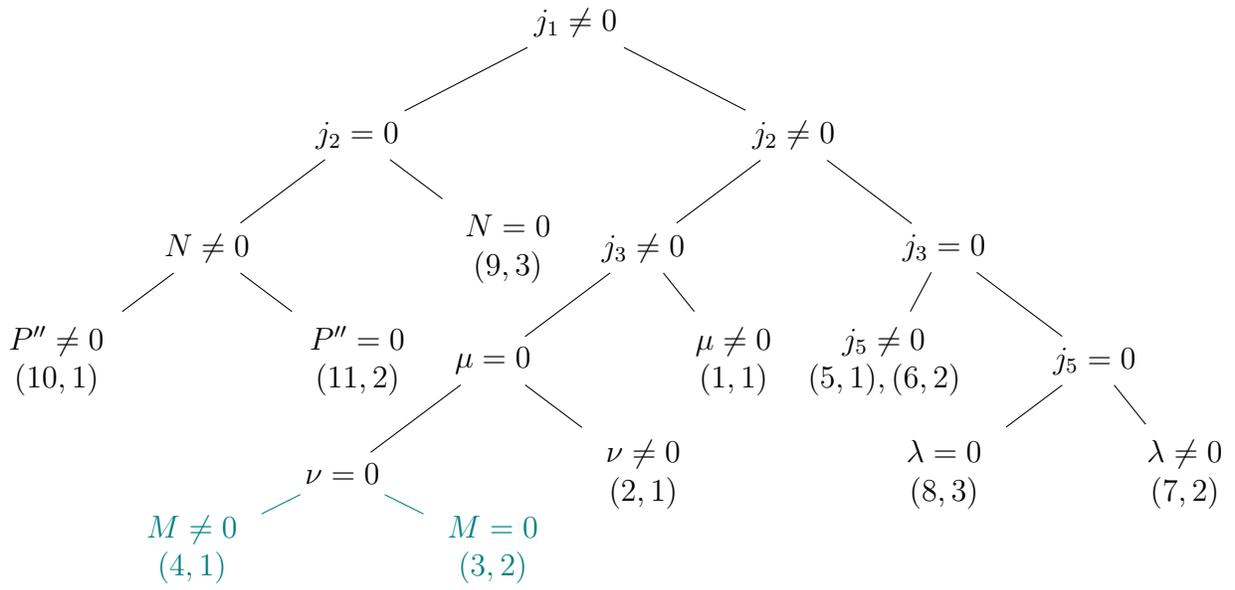
\begin{figure}[h!]
\centering
\hspace*{-5mm}
\begin{tikzpicture}
[
    level 1/.style = {sibling distance = 8cm},
    level 2/.style = {sibling distance = 4cm},
    level 3/.style = {level distance = 3cm}
    level 4/.style = {sibling distance = 1cm},
    level 5/.style = {teal, level distance = 1cm}
    ]
\node {$j_1\neq 0$}
    child {node [xshift = +1.1cm] {$j_2=0$}
    child {node {$N\neq 0$}
    child {node { \nodeii{P^{\prime\prime}\neq 0}{10, 1} }}
    child {node { \nodeii{P^{\prime\prime}=0}{11, 2} }}
    }
    child {node {  \nodeii{N=0}{9, 3} }}
    edge from parent }
    child {node [xshift = -1.1cm] {$j_2\neq 0$}
    child {node {$j_3\neq 0$}
    child {node {$\mu= 0$}
    child {node {$\nu= 0$}
    child {node {  \nodeii{M\neq 0}{4, 1} }}
    child {node { \nodeii{M=0}{3, 2} }}
    }
    child {node { \nodeii{\nu \neq 0}{2, 1} }}
    }
    child {node [xshift = -.8cm] { \nodeii{\mu\neq 0}{1, 1} }}}
    child {node {$j_3= 0$}
    child {node [xshift = +1.2cm] { \nodeiiarb{j_5\neq 0}{(5, 1), (6, 2)} }}
    child {node {$j_5=0$}
    child {node { \nodeii{\lambda= 0}{8, 3} }}
    child {node [xshift = -.8cm] {\nodeii{\lambda \neq 0}{7, 2}}}
    }}};

\end{tikzpicture}
\caption{Tree of the study for $j_1\neq 0$, $b_{01}^2+b_{02}^2\neq 0$ and $S\neq const$}\label{fig:j_1neq_0}
\end{figure}

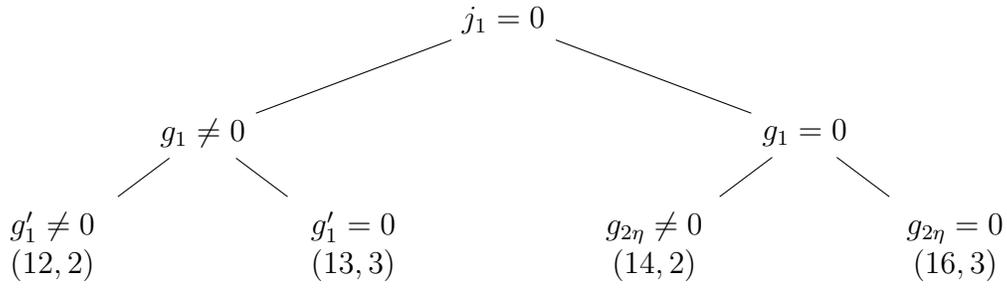
\begin{figure}[h!]
\centering
\hspace*{-5mm}
\begin{tikzpicture}
[
    level 1/.style = {sibling distance = 8cm},
    level 2/.style = {sibling distance = 4cm},
    level 3/.style = {level distance = 3cm}
    level 4/.style = {sibling distance = 1cm},
    level 5/.style = {teal, level distance = 1cm}
    ]

\node {$j_1= 0$}
    child {node {$g_1\neq 0$}
    child {node {\nodeii{g_1^\prime \neq 0}{12, 2} }}
    child {node { \nodeii{g_1^\prime =0}{13, 3} }}
    edge from parent }
    child {node {$g_1= 0$}
    child {node{\nodeii{g_{2\eta}\neq 0}{14, 2} }}
    child {node {\nodeii{g_{2\eta}= 0}{16, 3} }}};

\end{tikzpicture}
\caption{Tree of the study for $j_1= 0$, $b_{01}^2+b_{02}^2\neq 0$ and $S\neq const$}\label{fig:j_1eq_0}
\end{figure}

\begin{figure}[h!]
\centering
\hspace*{-5mm}
\begin{tikzpicture}
[
    level 1/.style = {sibling distance = 8cm},
    level 2/.style = {sibling distance = 4cm},
    level 3/.style = {level distance = 3cm}
    level 4/.style = {sibling distance = 1cm},
    level 5/.style = {teal, level distance = 1cm}
    ]

\node {$S= const$}
    child {node {$b_{01}^2+ b_{02}^2\neq 0$}
    child {node {\nodeii{\psi_{\eta\eta}\neq 0}{22, 4} }}
    child {node {\nodeii{\psi_{\eta\eta}=0}{23, 3} }}
    edge from parent }
    child {node {$b_{01}^2+ b_{02}^2= 0$}
    child {node {\nodeii{b_{03\eta}\neq 0}{24, 3 } }}
    child {node {\nodeii{b_{03}=const}{25, 2 } }}
    };

\end{tikzpicture}
\caption{Tree of the study for isentropic flows $S=const$.}\label{fig:s_const}
\end{figure}
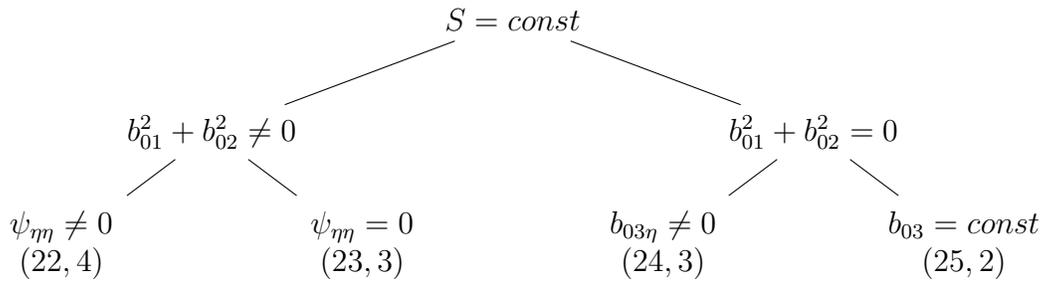

The transition to Lagrangian coordinates allows
integrating four equations of magnetogasdynamics of an ideal perfect
polytropic gas: the entropy $S(\xi,\eta)$ and the functions associated
with the magnetic field $(b_{01}(\xi,\eta),b_{02}(\xi,\eta),b_{03}(\xi,\eta))$
are arbitrary functions of the integration. This leads to complications
in the study of group classification: consideration of the many possibilities
of these functions. The analysis presented in this article gives a
complete investigation of all these possibilities. Figures \ref{fig:j_1neq_0}-\ref{fig:b1b2_eq_0}
provide the trees of the study of nonisentropic cases, where $(i,j)$
means the following: $i$ is the number of the extension of the kernel
of admitted Lie algebras (\ref{eq:kernel}), $j$ is the number of
the generators $X_{k+8}^{(i)},\ (k=1,2,...,j)$ in $i$th extension.
Figure \ref{fig:s_const} presents the tree of the study for isentropic
flows. The Lie algebras corresponding to the extensions $i$ ($i=1,2,...,19$)
are finite dimensional, the Lie algebras corresponding to other extensions
are infinite dimensional.

As mentioned above, finding an admitted Lie group is one of the first
and necessary steps in application of the group analysis method for
constructing invariant and partially invariant solutions. Because
the equations (\ref{eq:2Dequations}) are variational, the symmetries
found can also be used to derive conservation laws using Noether's
theorem. The wide variety of these symmetries allows us to expect
the derivation of new conservation laws. The search for invariant
solutions, as well as the derivation of conservation laws, are the
subject of further applications of the symmetries obtained in the
present work.

\section*{Acknowledgements}

The research was supported by Russian Science Foundation Grant No.~18-11-00238
`Hydro\-dynamics-type equations: symmetries, conservation laws, invariant
difference schemes'. E.I.K. acknowledges Suranaree University of Technology~(SUT)
and Thailand Science Research and Innovation~(TSRI) for Full-time
Doctoral Researcher Fellowship~(Full-time61/15/2021).

\vfill

\begin{figure}[h!]
\centering
\hspace*{-5mm}
\begin{tikzpicture}
[
    level 1/.style = {sibling distance = 8cm},
    level 2/.style = {sibling distance = 4cm},
    level 3/.style = {level distance = 3cm}
    level 4/.style = {sibling distance = 1cm},
    level 5/.style = {teal, level distance = 1cm}
    ]

\node {$b_{01}=b_{02}= 0$}
    child {node {$f_2\neq 0$}
    child {node {\nodeii{f_3 \neq const}{17, 1 } }}
    child {node {$f_3 = const$}
    child {node {\nodeii{f_4 = const}{18, 2 } }}
    child {node {\nodeii{f_4 \neq const}{19, 1 } }}
    }
    edge from parent }
    child {node {$f_2= 0$}
    child {node {\nodeii{f_5= const}{21, 3 } }}
    child {node {\nodeii{f_5 \neq const}{20, 2 } }}
    };

\end{tikzpicture}
\caption{Tree of the study for $b_{01}= 0$, $b_{02}=0$ and $S\neq const$}\label{fig:b1b2_eq_0}
\end{figure}
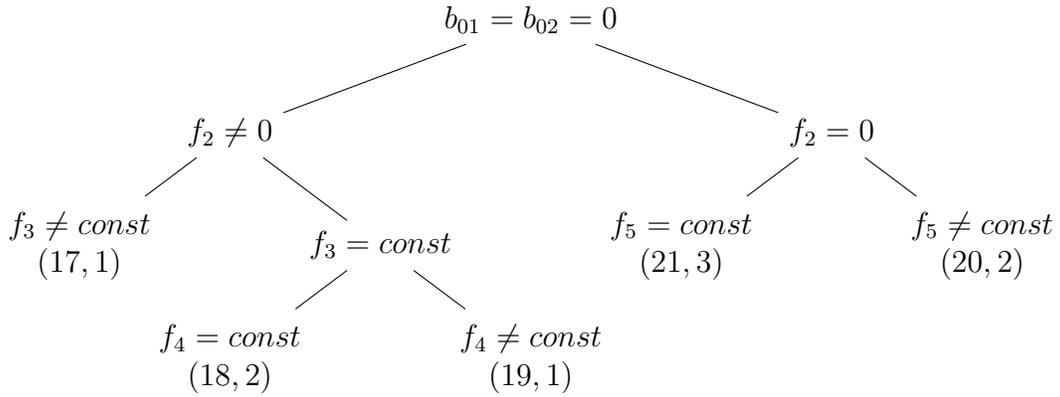

\pagebreak
\vspace*{+20mm}


\end{document}